\documentclass[a4paper,10pt]{article}
\usepackage[utf8]{inputenc}

\usepackage{geometry}
\usepackage[english]{babel}
\usepackage{lineno}
\usepackage{float}
\usepackage{amssymb}
\usepackage{amsmath, bm}
\usepackage{latexsym}
\usepackage{amsthm}
\usepackage{fancyhdr}
\usepackage{graphicx}
\usepackage{newlfont}
\usepackage{textcomp}
\usepackage{numcompress}
\usepackage{caption}
\usepackage{booktabs}
\newtheorem{remark}{Remark}

\usepackage{hyperref}
\usepackage{color}

\newcommand{\giachi}[1]{\textcolor{black}{{#1}}}

\title{Monolithic coupling of implicit material point method with finite element method}
\author{%
  Eugenio Aulisa \footnote{Department of Mathematics and Statistics, Texas Tech University}
  \and Giacomo Capodaglio\footnote{Department of Scientific Computing, Florida State University}%
  }
\date{}

\begin{document}
 
 \maketitle

\begin{abstract}
A monolithic coupling between the material point method (MPM) and the finite element method (FEM) is presented.
The MPM formulation described is implicit, and the exchange of information between particles and background grid is minimized. \giachi{The reduced information transfer from the particles to the grid improves} the stability of the method. 
Once the residual is assembled, the system matrix is obtained by means of automatic differentiation. In such a way, no explicit computation is required and the implementation is considerably simplified.
When MPM is coupled with FEM, the MPM background grid is attached to the FEM body and the coupling is monolithic.
With this strategy, no MPM particle can penetrate a FEM element, and the need for computationally expensive contact search algorithms used by existing coupling procedures is eliminated.
The coupled system can be assembled with a single assembly procedure carried out element by element in a FEM fashion.
Numerical results are reported to display the performances and advantages of the methods here discussed. 
\end{abstract}
 
%  \begin{keyword}
%  Implicit Material Point Method \sep Finite Element Method \sep Monolithic Coupling \sep Automatic differentiation \sep MPM-FEM Coupling.
% %\MSC[2010] 00-01\sep  99-00
% \end{keyword}

\section{Introduction}
The material point method (MPM) is a numerical method for problems that involve large deformations and/or history dependent materials.
It has been originally introduced in \cite{sulsky1994particle, sulsky1995application, sulsky1996axisymmetric} by Sulsky et al. as an extension to solid mechanics of the hydrodynamics FLIP method \cite{brackbill1986flip}.
The MPM has been applied to simulate numerous problems involving, for instance: membranes \cite{ionescu2005computational, zhou1999simulation}, granular materials \cite{wikeckowski1999particle, bardenhagen2000material}, sea ice modeling \cite{tran2015anisotropic}, explosions \cite{guilkey2007eulerian}, free surface flows \cite{zhang2017incompressible}, snow modeling \cite{stomakhin2013material}, viscoelastic fluids \cite{ram2015material}, phase-change \cite{stomakhin2014augmented}, fluid-structure interaction \cite{york2000fluid, gilmanov2008hybrid, hamad2017interaction}, and dynamic crack propagation \cite{guo2006three}.
The MPM exploits the advantages of both the Eulerian and the Lagrangian approaches. The body is discretized with a set of particles (or material points) and is positioned on a background grid, where the momentum equation is solved.
The background grid may be chosen to be a finite element grid, because interpolating functions are used to transfer information from the particles to the grid and back. In the classical MPM, particles information is transferred to the background grid, where the momentum equation is solved. The new information obtained from the solution of the momentum equation at the grid nodes is used to update the values of displacement, velocity and acceleration at the particles. Then, the background grid is reset to its initial position. With such a method, the drawbacks of using a purely Lagrangian approach are avoided, because entanglement of the background grid is prevented by resetting it to its initial state after every time step. Moreover, the numerical dissipation associated with an Eulerian approach is eliminated, due to the absence of the convection term. 

Historically, the MPM has been conceived as an explicit method.
Compared to the extensive amount of explicit MPM formulations available in the literature, only a few efforts have been made so far to set the analysis in an implicit framework \cite{ stomakhin2014augmented, guilkey2003implicit, wang2016development, charlton2017igimp, cummins2002implicit, sulsky2004implicit, love2006unconditionally, love2006energy, beuth2008large}.
An implicit approach is desirable because it guarantees much larger time steps compared to what can be obtained with explicit methods, \giachi{but more importantly the implicit formulation facilitates the monolithic coupling of the MPM with the FEM, which is the ultimate goal of the present work.}
% In this work, a monolithic coupling between an implicit material point method and the finite element method is presented.
As other existing implicit MPM formulations, here the assembly procedure for the problem on the background grid is carried out in a finite element fashion. The only difference between a standard FEM assembly is that, for elements of the background grid that enclose MPM particles, the quadrature points used for numerical integration are not the Gauss points as in standard FEM, but rather
the particles themselves. 
The implicit MPM formulated here differs from existing implicit approaches \cite{guilkey2003implicit, wang2016development, sulsky2004implicit, love2006unconditionally},
because the exchange of information between particles and grid is minimized
by avoiding {\it unnecessary} projections from the particles to the background grid nodes.
In turn, as it is shown later, the inertial contribution is evaluated directly at the particles rather than at the grid nodes.
This approach is consistent with an assembly procedure where the particles are used in place of the Gauss points.
A drawback of using particles as quadrature points consists of instabilities that may arise due to inaccurate numerical integration, which may occur when an element of the background grid does not host enough particles.
To overcome this issue, a {\it{soft stiffness}} matrix is added to the MPM system matrix as in \cite{wang2016development}.
In the present work, the magnitude of the soft stiffness contribution on a given element depends on how many particles the element itself and its neighbors are hosting. This is measured with appropriate flags whose determination procedure is discussed in Section \ref{numAlg}.
Once these flags are obtained, scaling factors associated with them are computed. The scaling factors are involved in the assembly procedure of the soft stiffness matrix and determine the weight of its contribution on a given element.
Clearly, it is necessary to strike a balance between the need for numerical stability, given by larger soft stiffness contributions, and the desire for accuracy, which improves as the soft contributions tend to zero. 
% An appropriate tuning of the scaling factors has to be done depending on the specific application considered, examples will be shown in Section \ref{numRes}.
% \giachi{As a general heuristic rule, the scaling factors are proportional to the shear modulus and lie in the interval $[10^{-8},10^{-2}]$ to ensure stability of the method. Examples are shown in Section \ref{numRes}.}
To further improve numerical stability \giachi{and simplify the implementation}, the background grid is divided in active and inactive background grid.
The former is composed of all the elements of the background grid that host at least one particle,
whereas elements that do not host any particle are collected in the inactive background grid.
The active background grid is what is actually used at a given instant of time to interpolate the particle instances from the grid.
Contributions from nodes that belong exclusively to elements of the inactive background grid are assembled in a separate way using a lumped mass matrix. Details on this procedure are given in Section \ref{numAlg}.
\giachi{Taking into account the inactive background grid contributions, we can avoid resizing the global system of equations every time a particle moves to a different element. Moreover, the solution corresponding to the lumped mass matrix block is inexpensive (one Jacobi iteration), because it corresponds to the inverse of a diagonal matrix.}
Another feature of the proposed formulation resides on the use of implicit differentiation for the computation of the system matrix.
After the residual is assembled, the system matrix is obtained differentiating the residual vector with the library Adept \cite{hogan2014fast}.
With such a tool, the system matrix does not have to be computed explicitly as in existing implicit MPM strategies, and
the implementation is strongly simplified.

For what concerns the coupling of MPM and FEM, only very few studies are available in the literature \cite{lian2011coupling, chen2015improved, lian2012coupling, lian2012adaptive}.
For simplicity, it is assumed that the interaction to be simulated is that of two solid bodies:
one undergoing large deformations, modeled with MPM and one subject to small deformations, modeled with FEM.
Such a choice is justified by the respective advantages of the two discretization methods when applied to different deformation regimes.
The assembly procedure is carried out according to a finite element strategy and automatic differentiation is used to obtain the coupled system matrix.
The proposed monolithic algorithm solves simultaneously for the MPM and FEM unknowns in a unique solver, so that the two solid bodies
are treated as a single continuum. 
The stress balance and the kinematic conditions across the MPM-FEM interface are automatically satisfied, resulting in a coupling that is accurate, robust, and stable. 
To allow a simultaneous solution of MPM and FEM unknowns, a finite element grid attached to the FEM body is used as background grid for the MPM. With such a feature, the interface between the MPM body and the FEM body is automatically tracked. No MPM particle can penetrate a FEM element, and a great advantage in terms of computational time is obtained, since time consuming contact search algorithms like the ones adopted in \cite{lian2011coupling, chen2015improved} are eliminated.
All the algorithms proposed in this work are implemented in the in-house finite element library FEMuS \cite{femus-web-page}.

The paper is structured as follows: in Section \ref{domainConf}, the domain configurations later employed in the mathematical formulation are introduced. In Section \ref{implicitMPM}, a detailed description of the implicit MPM formulation proposed is provided. The new features of our formulation and the complete numerical algorithm are presented and discussed.
In Section \ref{couplingMPMFEM}, the monolithic coupling between MPM and FEM is addressed, and the characteristics of the coupled problem are laid out. \giachi{In Section \ref{receding}, an algorithm for the receding phase of the contact interaction is described. This algorithm is necessary in order to prevent {\emph{sticky}} phenomena induced by our monolithic formulation for large values of the Young's modulus.} The paper is concluded with Section \ref{numRes}, where numerical results are reported to illustrate the performances of the proposed algorithms.

\section{Domain Configurations}\label{domainConf}
In this section, the domain configurations used in the mathematical formulation are discussed.
A general description is given first, followed by the identification of what will be referred to as the MPM domain and the FEM domain.
The problem of the reset of the MPM background at the beginning of every time step is addressed for both the uncoupled and the coupled case.

\begin{figure}[t]
\centering
\includegraphics[scale=0.5]{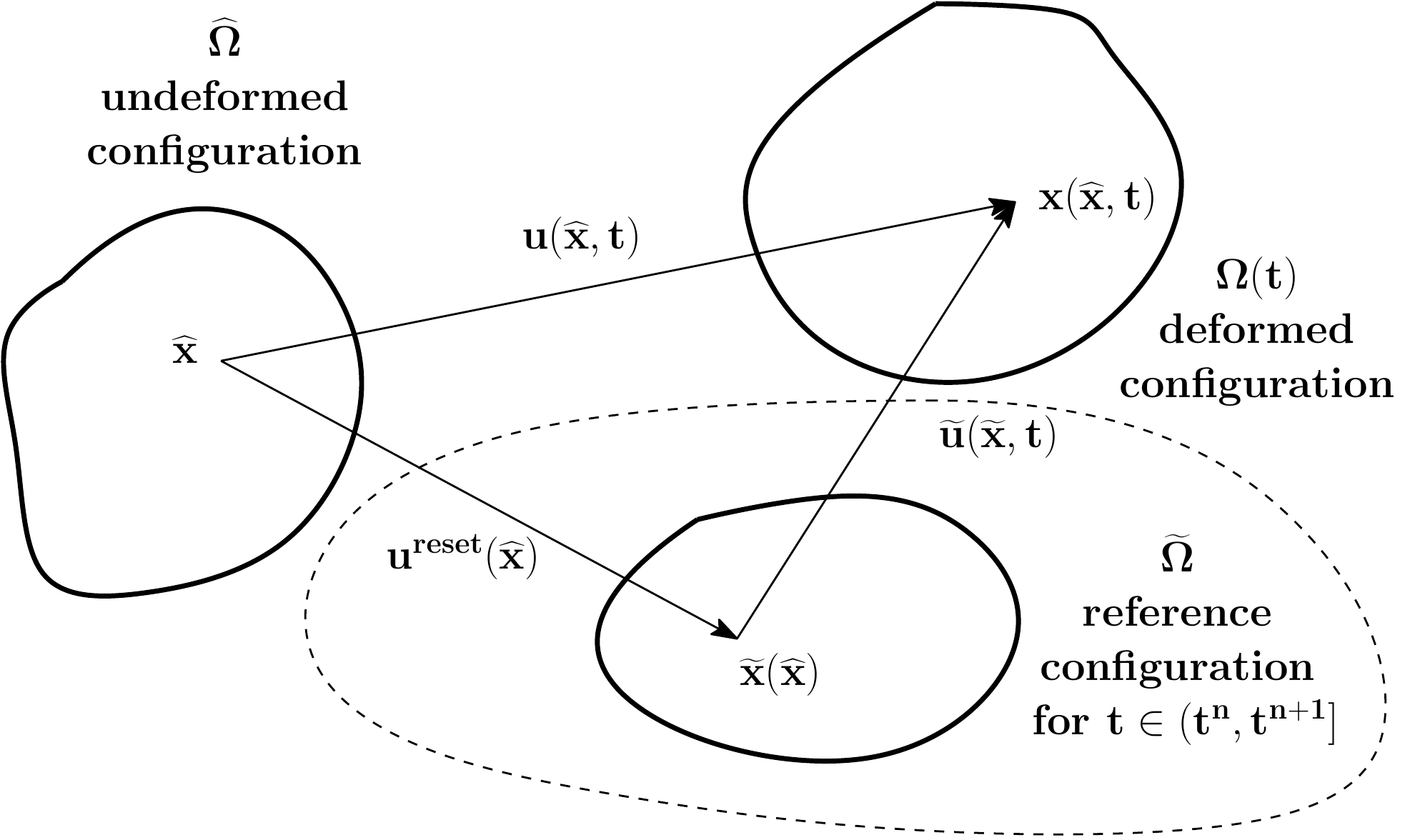}
\caption{Schematics for the domain configurations.}\label{domainSchematics}
\end{figure}
\subsection{Three general domain configurations}
Three different domain configurations are used in the mathematical formulation. 
The reason why the configurations are three is made explicit in the following sections. 
Let $T>0$ and let $\Omega$ be a given open and bounded domain such that, for all $t \in [0,T]$, $\Omega(t) \subset \mathbb{R}^d$. Then, the {\it{undeformed configuration}} of $\Omega$
is indicated with the notation $\widehat\Omega$ and it represents the position of the domain in its original 
state at time $t=0$. For the {\it{deformed configuration}} $\Omega(t)$, no superscript is used, and the notation $\Omega(t)$ 
just refers to the domain in its deformed configuration at time $t>0$. This configuration is obtained 
from the undeformed configuration with a translation. 
Specifically, if $\widehat{\bm x}$ represents any point of $\widehat{\Omega}$, then 
\begin{align}
 {\bm x} (\widehat{\bm x}, t) = \widehat  {\bm x} + \bm { u } (\widehat  {\bm x},t), \quad  \text{ for }  t > 0,
\end{align}
where $\bm { u } (\widehat  {\bm x},t)$ is the displacement field.
The third configuration to be introduced is the {\it{reference configuration}}, denoted by
$\widetilde \Omega$. We remark that, given $n\geq0$, such a configuration does not change for any $t\in(t^n,t^{n+1}]$.
The reference configuration can also be obtained from the undeformed configuration with a translation, as follows
\begin{align}\label{relationReset}
 \widetilde {\bm x} (\widehat {\bm x} ) = \widehat  {\bm x} + {\bm u}^{reset} (\widehat {\bm x}), \quad\;  \text{ for }  t  \in (t^n, t^{n+1}],
\end{align}
where the field $\bm{u}^{reset}$ is referred to as the {\it{reset displacement field}}
associated with the given time interval $(t^n,t^{n+1}]$.
The definition of such a field is given in the next section.
Any point in the deformed configuration can be also obtained from the reference configuration as
\begin{align}
 {\bm x} (\widetilde{\bm x},t) = \widetilde {\bm x} + \widetilde{\bm  u} (\widetilde{\bm x},t), \quad  \text{ for }  t  \in (t^n, t^{n+1}],
\end{align}
where the field $\widetilde{\bm{u}}$ is the displacement field with respect to the reference configuration, which satisfies
$$  \widetilde{\bm  u} (\widetilde{\bm x}(\widehat{\bm x}),t) = \bm { u } (\widehat  {\bm x},t) - {\bm u}^{reset} (\widehat {\bm x}), \quad  \text{ for }  t  \in (t^n, t^{n+1}].$$
Schematics for the three domain configurations can be found in Figure \ref{domainSchematics}.

The general considerations about domain configurations are applied to the MPM and FEM domains as explained next.

\subsection{The MPM and the FEM domains}\label{MPMFEMdomains}
Let $T>0$ and $[0,T] \subset \mathbb{R}$. For all $t \in [0,T]$, let $\Omega_{mpm}(t)$ and $\Omega_b(t)$ be open and bounded subsets 
of $\mathbb{R}^d$. Assume that $\Omega_{mpm}(t) \subset \Omega_b(t)$ for all $t \in [0,T]$.
From now on, $\Omega_{mpm}(t)$ represents the body discretized with the MPM particles. The set $\Omega_b(t)$ 
is discretized with a regular finite element grid $\mathcal{T}_b$ \cite{Ciarlet:2002:FEM:581834, brenner2007mathematical}, 
used as a background grid for the MPM body.
While $\Omega_{mpm}(t)$ may be subject to roto-translation and large deformations due to the movement of the MPM body, 
$\Omega_b(t)$ undergoes only limited deformations, because it follows the MPM body motion only in the time interval $(t^n,t^{n+1}]$. 
At the beginning of each time step, $\Omega_b(t)$ is 
reset to its reference configuration $\widetilde{\Omega}_b$, as it will be explained in the next section.
 \begin{figure}[t]
\centering
\includegraphics[scale=0.8]{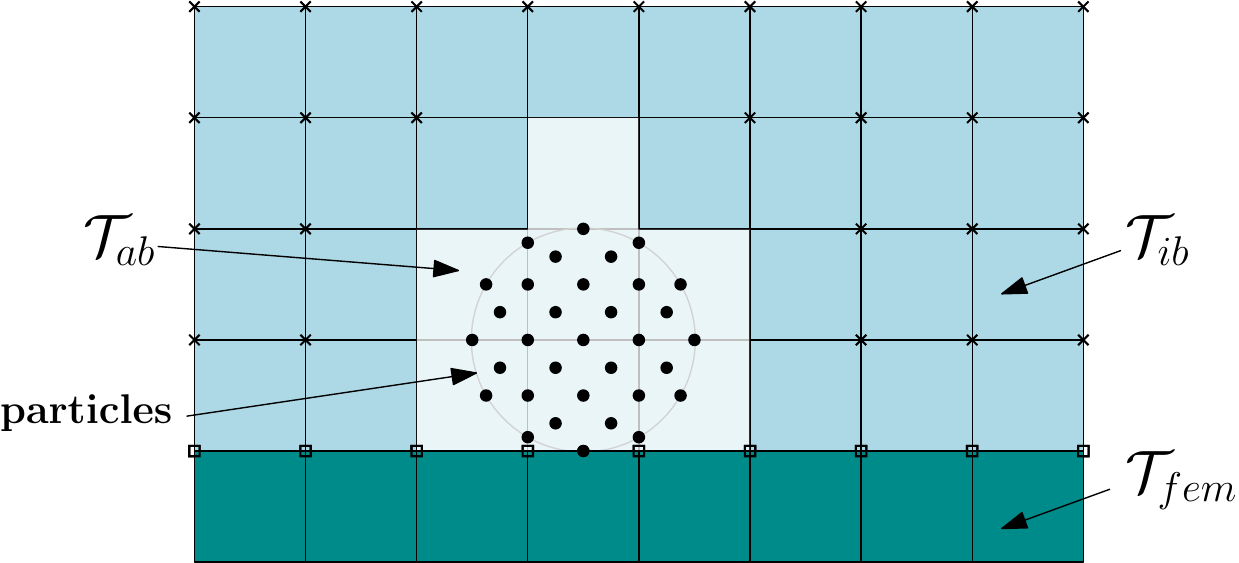}
\caption{\giachi{Schematics of triangulations and domains used in the MPM-FEM coupled problem. The nodes marked with a cross belong exclusively to the inactive background grid $\mathcal{T}_{ib}$, while those marked with a box are the nodes at the interface between the MPM background grid $\mathcal{T}_b$ and the FEM grid $\mathcal{T}_{fem}$.}}\label{figsec2}
\end{figure}
Denoting by $\mathcal{E}_p$ the element of the background grid that hosts a given particle $p$, 
we introduce 
\begin{align}\label{omegaActive}
\Omega_{ab}(t) \equiv \bigcup \limits_{\mathcal{E} \in \mathcal{T}_{ab}} \mathcal{E} = \bigcup \limits_{p=1}^{N_p} \mathcal{E}_p, 
\end{align}
where $t \in (t^n,t^{n+1}]$, and $\mathcal{T}_{ab} \subset \mathcal{T}_b$ is the active background grid, composed of all elements $\mathcal{E}$ of $\mathcal{T}_b$ that host at least one particle $p$. 
Similarly, we define $\mathcal{T}_{ib} \equiv \mathcal{T}_b \setminus \mathcal{T}_{ab}$ to be the inactive background grid, composed of all elements of $\mathcal{T}_b$ that do not enclose any particle.
Note that the time dependence of $\Omega_{ab}(t)$ is due to the time dependence of $\mathcal{E}_p$ (for given $p$) in the time interval $(t^n,t^{n+1}]$. 
Moreover, because the background grid is reset at the beginning of each time step, a particle $p$ may leave a particular element $\mathcal{E}_p$ 
and move into one of its neighboring elements, so the elements of $\mathcal{T}_{ab}$ and $\mathcal{T}_{ib}$ 
may also change at each time step. \giachi{Schematics of the grids here described are shown in Figure \ref{figsec2}.}
By definition, $\Omega_{ab}(t)$ has the property that $\Omega_{mpm}(t) \subseteq \Omega_{ab}(t) \subseteq \Omega_b(t)$.
The set $\Omega_{ab}(t)$, associated with the MPM active background grid, is what we refer to as the {\it{MPM domain}}. 
Such a choice is motivated by the fact that the MPM momentum equation is solved only on the active background grid, 
and its elements consequently undergo deformations that modify their reference configuration.

Concerning the FEM, since the triangulation entirely covers the body to discretize, 
the open and bounded set $\Omega_{fem}(t) \subset \mathbb{R}^d$ is used to characterize both the body discretized with FEM and the FEM domain. 
The interfaces between the FEM domain and the background grid $\Omega_b(t)$ and between 
the FEM domain and the active background grid $\Omega_{ab}(t)$
that arise during the coupling procedure are indicated with the letters $\Gamma$ and $\Gamma_a$, respectively.

\subsection{Definition of the reset displacement field}

In the classical MPM method, where no coupling between MPM and FEM occurs, at the beginning of each time step interval $(t^n, t^{n+1}]$, 
the set $\Omega_b(t)$ is reset to its undeformed configuration, $\widehat{\Omega}_b$, to avoid mesh entanglement. 
Namely, for any $t \in (t^n, t^{n+1}]$, if $I$ refers to any node of the grid, the grid displacement $\bm{u}_{I}$ associated to the node $I$ is such that 
\begin{align}\label{classicalMPM}
\lim\limits_{\; t\rightarrow {(t^n)}^+}\bm{u}_I(t) =\bm 0.
\end{align}
Moreover, in general 
\begin{align}
 \lim\limits_{\; t\rightarrow {(t^n)}^-}\bm{u}_I(t) = \bm{u}_I(t^n)\ne \bm 0,
\end{align}
and as a result $\bm{u}_{I}(t)$ is discontinuous in time. 
On the other hand, in the classical uncoupled FEM method, the grid follows the solid body deformation and its displacement 
is continuous in time
\begin{align}
 \lim\limits_{\; t\rightarrow {(t^n)}^-}\bm{u}_I(t) = \bm{u}_I(t^n) = \lim\limits_{\; t\rightarrow {(t^n)}^+}\bm{u}_I(t).
\end{align}
With a monolithic coupling approach, there is a unique displacement field defined on the MPM background grid and on the finite element grid discretizing the FEM body. Hence, when the coupling between MPM and FEM is considered, the displacement at the grid points of $\mathcal{T}_b$ that lie on the interface $\Gamma$ between $\Omega_b(t)$ and the FEM body cannot be reset to zero. 
This implies that at every instant of time, the deformed configuration of the MPM domain is composed of elements 
that are deformed, not only because of the motion of the MPM body, but also because of the FEM body motion. 

The requirement of resetting the MPM background grid to its undeformed configuration at the beginning of every time step 
is what motivates the introduction of the reset displacement field to be used for the coupled problem.
More specifically, when MPM is coupled with FEM, the background grid has to fulfill the following constraints:
in the FEM domain and on the interface $\Gamma$, it has to follow the solid deformation;
in the MPM domain, in order to enforce domain continuity, 
it has to follow the deformation of $\Gamma$, coming from the FEM domain.
Moreover, to avoid mesh entanglement, it is desirable to reset the background grid to a configuration similar to the undeformed configuration on all MPM nodes that are sufficiently far from $\Gamma$.

Because in the applications described in this paper the FEM solid body is subject to small deformations,
the background grid constraints are satisfied with the following definition of the reset displacement field $\bm{u}_{I}^{reset}$.
Let $\mathcal{I}$ be the set of grid points of $\mathcal{T} = \mathcal{T}_b \cup \mathcal{T}_{fem}$, where $\mathcal{T}_{fem}$ denotes the finite element grid used to discretize the FEM body. 
For all $t \in (t^n, t^{n+1}]$,  the reset displacement field satisfies \giachi{
\begin{equation}\label{CE1a}
 \begin{aligned} 
 \begin{cases}
 \bm{u}_{I}^{reset} = \lim\limits_{\; t\rightarrow {(t^n)}^-}\bm{u}_I(t) &, \mbox{ if} \; \bm x_I \in \Omega_{fem}(t) \,\,\,\\
 \nabla \cdot \left(\nabla \bm{u^{reset}}_{I} + (\nabla \bm{u^{reset}}_{I})^{T} \right)  =\bm{0} &, \mbox{ if} \; \bm x_I \in \Omega_{b}(t) \setminus \Gamma  
 \end{cases}
 \end{aligned}
 \end{equation}}
where $\bm{x}_I$ denotes the position of node $I \in \mathcal{I}$.
When MPM is coupled with FEM, Eq. \eqref{classicalMPM} is replaced with 
\begin{equation}\label{coupledMPMFEM}
\lim\limits_{\; t\rightarrow {(t^n)}^+}\bm{u}_I(t) =\bm{u}_{I}^{reset}. 
\end{equation}
By making this choice, %only one layer of MPM elements (the ones whose boundary partially overlaps with $\Gamma$) may have 
%$ \lim\limits_{\; t\rightarrow {(t^n)}^+}\bm{u}_I(t) \ne \bm 0 $. Moreover, 
continuity of the domain is preserved across $\Gamma$, 
and the grid $\mathcal{T}_b$ follows the solid body deformation on $\Omega_{fem}(t)$, resulting in a continuous displacement field $\bm{u}_I(t)$.
As a consequence of Eq. \eqref{coupledMPMFEM}, at the beginning of each time step, the deformed configuration of the background grid coincides with its reference configuration, although it is in general different from its undeformed configuration $\widehat{\Omega}_{b}$.
\giachi{This is more perceptible on those elements close to $\Gamma$ that may deform more to follow the FEM body motion. 
If the FEM body deformation is small, than the deformation of the background grid will also be small.}
To fulfill the classical MPM requirement in Eq. \eqref{classicalMPM}, a new MPM displacement field based on the reference 
background grid configuration is defined as
\begin{align}\label{nodalUTilde}
\widetilde{\bm{u}}_{I}(t) = \bm{u}_{I}(t) - \, \bm{u}_{I}^{reset}.
\end{align}
By Eq. \eqref{coupledMPMFEM}, this new field satisfies  
 $$\lim\limits_{\; t\rightarrow {(t^n)}^+}\widetilde{\bm{u}}_{I}(t) =\bm 0,$$
as in the classical uncoupled  MPM method.
The reference configuration is the configuration of the MPM domain associated with the field $\widetilde{\bm{u}}_{I}(t)$.

\section{Implicit Material Point Algorithm}\label{implicitMPM}
The mathematical formulation of the proposed implicit material point and the complete numerical algorithm are described in this section. 

\subsection{Mathematical Formulation}
For ease of notation, during the rest of this paper, the time dependence of $\Omega_{mpm}$, $\Omega_{ab}$, and of the fields involved in the formulation is not made explicit. The mass conservation is automatically satisfied by the material point method \cite{zhang2016material}, therefore the governing equations for the MPM body consist of momentum conservation with appropriate boundary and initial conditions.
For simplicity, we consider zero boundary conditions for displacement and normal stress
\begin{equation} \label{momentum}
\begin{aligned} 
& \rho \, \ddot{\bm{u}} - \nabla \cdot \bm{\sigma}  = \rho \, \bm{b}, & \mbox{in }  \Omega_{mpm} \times [0,T], \\
& \bm{\sigma}\cdot \bm{n}   = 0 , & \mbox{on }  {\partial \Omega_{mpm,trac}}\times [0,T],\\
& {\bm{u}}  = 0 , & \mbox{on }  {\partial \Omega_{mpm,disp}}\times [0,T],  \\
& {\bm{u}}(\bm{x},0)  = {\bm{u}}_0, \quad \dot{\bm{u}}(\bm{x},0) = \dot{\bm{u}}_0 & \mbox{in } \Omega_{mpm} .
\end{aligned}
\end{equation}
\giachi{The case of Robin and mixed boundary conditions is not addressed at present.}
In Eq.\eqref{momentum}, $\bm{u}$ represents the displacement,  $\bm{\sigma}$ is the Cauchy stress tensor, $\rho$ is the MPM body density, $\bm{b}$ is the body force per unit mass, $\dot{\bm{u}}$ is the velocity, $\ddot{\bm{u}}$ is the acceleration and $\bm{n}$ is the unit outward normal of the boundary $\partial \Omega_{mpm,trac}$, that represents the traction boundary. Similarly, $\partial \Omega_{mpm,disp}$ represents the portion of the boundary where displacement boundary conditions are prescribed. 
Constitutive equations  for the Cauchy stress need to be added to complete the above set of equations. We describe the MPM solid bodies as Neo-Hookean materials, with the Cauchy stress tensor given by \cite{ogden1997non}
\begin{align}\label{CauchyNeoHookean}
\bm{\sigma} = \lambda  \dfrac{\log(J)}{J} \, \bm{I} + \dfrac{\mu}{J} \, \Big( \bm{B} - \bm{I}\Big).
\end{align}
In Eq. \eqref{CauchyNeoHookean}, $\lambda$ is Lam\'{e}'s first parameter, $\bm{I}$ is the identity matrix, $\mu$ is the shear modulus, $\bm{B} = \bm{F} \, \bm{F}^T$ is the left Cauchy-Green strain tensor, $\bm{F}$ is the deformation gradient and
$J=\det(\bm{F})$.
Let $\Delta U = \{ \bm{\delta u} \in \bm{H}^1(\Omega_{mpm}) \,\, | \,\, \bm{\delta u}|_{\partial \Omega_{mpm,disp}} = 0  \}$.
The virtual displacements are chosen as test functions for the weak formulation.
The weak formulation  of the momentum balance reads
\begin{align}\label{weak}
\int_{\Omega_{mpm}} \rho \, \Big(\ddot{u_i} \, \delta u_i \,  +  \, \sum_{j=1}^d \sigma^{s}_{ij} \, \dfrac{\partial\delta u_{i}}{\partial x_j} \, -  \, b_i \, \delta u_i \Big) \, dV 
%- \int_{\partial \Omega_{trac}} \rho \, \widetilde{t}^{s}_i \, \delta u_i \, dA 
=0, \,\, \mbox{for} \,\,i=1, \ldots,d.
\end{align}
where $\sigma^s_{ij} = \sigma_{ij} / \rho(\bm{x},t)$ is the specific Cauchy stress.
%and $\widetilde{t}^s_i = \widetilde{t}_i / \rho$ is the specific traction.
The interested reader can consult \cite{zhang2016material} for more details on how to derive the above weak formulation.
The integrals involved in Eq. \eqref{weak} can be transformed in sums if the MPM density $\rho$ is approximated using the Dirac delta function $\delta$  as
\begin{align}\label{mpmDensity}
\rho(\bm{x},t) \approx \sum_{p=1}^{N_p} m_p \delta(\bm{x} - \bm{x}_p(t)), 
\end{align}
where, $\bm{x}_p$ denotes the position of particle $p$, and $N_p$ represents the total number of particles used to discretize $\Omega_{mpm}$.
Note that the number of particles $N_p$ is time independent, assuming that no particle is leaving $\Omega_b$.
Substituting Eq.\eqref{mpmDensity} in Eq.\eqref{weak} we obtain 
\begin{equation}
\begin{aligned}\label{discreteWeak}
 \sum_{p=1}^{N_{p}} m_p \Big[& \ddot{u}_i(\bm{x}_p) \delta u_i(\bm{x}_p) + \sum_{j=1}^d\sigma_{ij}^{s}(\bm{x}_p) \dfrac{\partial\delta u_{i}}{\partial x_j}(\bm{x}_p)  
 - b_i(\bm{x}_p)\delta u_i(\bm{x}_p) 
%- \widetilde{t}_i^{s}(\bm{x}_p) h^{-1} \delta u_i(\bm{x}_p) 
\Big]=0.
\end{aligned}
\end{equation}
In the above equation, $\sigma_{ij}^{s}(\bm{x}_p) = J(\bm{x}_p) \, \sigma_{ij}(\bm{x}_p) / \rho_0$, where $\rho_0$ is the density of the MPM body in the initial, undeformed configuration.
\begin{remark}
Since the MPM body is discretized with particles, the values of the fields at the particles can be identified with the particle fields.
For instance, $\ddot{u}_i(\bm{x}_p) = \ddot{u}_{i,p}$, where $\ddot{u}_{i,p}$ represents the $i$-th component of the acceleration of particle $p$.
\end{remark}
Considering the above remark, Eq. \eqref{discreteWeak} can be rewritten as
\begin{equation}
\begin{aligned}\label{discreteWeakParticles}
 \sum_{p=1}^{N_{p}} m_p \Big[& \ddot{u}_{i,p} \delta u_{i}(\bm{x}_p) + \sum_{j=1}^d\sigma_{ij,p}^{s} \dfrac{\partial\delta u_{i}}{\partial x_j}(\bm{x}_p) 
 - b_{i,p}\delta u_{i}(\bm{x}_p) 
%- \widetilde{t}_{i,p}^{s} h^{-1} \delta u_{i}(\bm{x}_p) 
\Big]=0.
\end{aligned}
\end{equation}
The weak formulation for the MPM equations in Eq. \ref{discreteWeakParticles} is solved numerically employing the active background grid $\mathcal{T}_{ab}$.
For any given particle $p$, an interpolation from the nodes of the element $\mathcal{E}_p \in \mathcal{T}_{ab}$ that encloses $p$ is performed. This step is part of the assembly procedure of the background grid system and is discussed within the complete numerical algorithm.

\subsection{Implicit Time Integration}
In a time dependent framework, the position, displacement, velocity and acceleration of a given particle are functions of time.
The Newmark-beta \giachi{integrator} is adopted for the numerical integration of $\ddot{u}_{i,p}$ in Eq. \eqref{discreteWeakParticles}. If $\bm{u}^{n+1}_{p}$ denotes the particle displacement at time $t^{n+1}$ and the same notation is also adopted  for particle velocity and acceleration, the method gives
 \begin{align}\label{Newmark1}
  \dot{\bm{u}}^{n+1}_{p} &= \dot{\bm{u}}^{n}_{p} + (1 - \gamma) \,\, \Delta t \,\, \ddot{\bm{u}}^{n}_{p} + \gamma \,\, \Delta t \,\, \ddot{\bm{u}}^{n+1}_{p} , 
  \end{align}
  \begin{align}\label{Newmark2}
  {\bm{u}}^{n+1}_{p} &= {\bm{u}}^{n}_{p} + \Delta t \,\, \dot{\bm{u}}^{n}_{p} + \frac{1}{2} \Delta t^2 \,(1-2 \beta) \ddot{\bm{u}}^{n}_{p} + \giachi{\Delta t^2} \beta \ddot{\bm{u}}^{n+1}_{p},
 \end{align}
for $\gamma \in [0,1]$, $\beta \in [0,0.5]$ and $\Delta t = t^{n+1} - t^{n}$.
It follows that,
\begin{align}\label{Newmark3}
 \ddot{\bm{u}}^{n+1}_{p} = \dfrac{1}{\beta \, \Delta t^2} \, \Big({\bm{u}}^{n+1}_{p} - {\bm{u}}^{n}_{p}\Big)
 -\dfrac{1}{\beta \Delta t} \,\, \dot{\bm{u}}^{n}_{p} - \dfrac{1-2\beta}{2 \beta}\ddot{\bm{u}}^{n}_{p}.
\end{align}
\subsection{The numerical algorithm}\label{numAlg}
The complete numerical algorithm for the implicit MPM scheme presented in this paper is described next.
The algorithm is divided into a series of stages and it is set in a parallel framework where different processes handle their own share of the background grid and particles.

\begin{itemize}

\item {\bf{Initialization of the MPM particles}}.
\newline
The first step consists of particle initialization.
We assume that the total number of particles $N_p$, the volume $V_p$ associated with each particle,
and the initial density $\rho_0$ of the undeformed MPM body are given.
With this information, the mass of each particle is computed as $m_p = \rho_0 \, V_p$.
The initial position $\bm{x}_p$ of each particle $p$ is also known.
The values of displacement $\bm{u}_p$ and velocity $\dot{\bm{u}}_p$ of the particles are initialized as in Eq.~\eqref{momentum}.
The acceleration $\ddot{\bm{u}}_p$ is set to zero, while the deformation gradient 
$\bm{F}_p$ at the particle is initialized to a given initial value $\bm{F}^0_p \equiv \bm{F}_p(0)$.
Once a particle has been placed on the FEM background grid, the element $\mathcal{E}_p$ on the background grid that hosts the particle is determined, using the point locating method described in \cite{capodaglio2017particle}. As already mentioned above, such an element may be different over time, so for a given $p$, $\mathcal{E}_p$ is actually a function of time. For ease of notation, the time dependence is not made explicit. 
While for each $p$ the value of $\mathcal{E}_p$ is known to all processes, the local coordinates $\bm{\xi}_p$ of $p$ are computed only by the process that owns it, so the other processes do not possess this information.
To obtain the local coordinates $\bm{\xi}_p$ in the FEM reference configuration, an inverse mapping obtained with Newton's method is employed. For more details on this procedure please see \cite{capodaglio2017particle}.\\

\item {\bf{Assembly and solution of the background grid system}}.
\newline
For this step, it is assumed that the particle instances at time $t^n$ are known and that those at time $t^{n+1}$ have to be determined. 
Let $\mathcal{I}_b=\{1,\dots,N_b\}$ be the set of grid points of $\mathcal{T}_b$. 
Let $\mathcal{I}_{ab}$ be the set of grid points associated with $\mathcal{T}_{ab}$ and $\mathcal{I}_{ib} = \mathcal{I}_b \setminus \mathcal{I}_{ab}$. The assembly procedure is carried out differently depending on whether a node of the background grid belongs to the active or inactive background grid. Both assembly procedures are described in detail in the next steps.
\begin{enumerate}
\item {\bf{Assembly of the active background grid contributions}}.
\newline
A loop on the particles is executed to assemble the MPM contributions.
Given a particle $p$ from the loop, the element $\mathcal{E}_p$ on the active background grid $\mathcal{T}_{ab}$ that is currently hosting the particle is extracted.
In a sense, this procedure is similar to a standard FEM procedure since the assembly is still performed in an element-by-element fashion. The differences are that now particles are used as quadrature points instead of Gauss points, and that the quadrature contributions on a given element are not computed sequentially, since particles that belong to the same element may not be all grouped together in the particle loop. 
\begin{remark}
 It is important to point out that, unlike other existing implicit MPM formulations \cite{guilkey2003implicit, wang2016development, sulsky2004implicit, love2006unconditionally}, our assembly procedure does not start with a mapping from the particles to the grid nodes.
The Newmark scheme in Eq.\eqref{Newmark3} is used directly on the particle instances instead of the grid instances as existing strategies do. For this reason, no mapping from the particles to the background grid is necessary.
\end{remark}
According to a FEM approach, the particle virtual displacement in Eq.\eqref{discreteWeakParticles} can be obtained through interpolation from the nodes of $\mathcal{E}_p$.
If $N_{\mathcal{E}_p}$ denotes the number of nodes of element $\mathcal{E}_p$,  we have $\delta u_{i,p} = \sum_{I_{\mathcal{E}_p}=1}^{N_{\mathcal{E}_p}} \delta u_{i,I_{\mathcal{E}_p}} \phi_{I_{\mathcal{E}_p}}(\bm{x}_p)$, where $\delta u_{i,I_{\mathcal{E}_p}}$ represents the nodal value of the virtual displacement at node $I_{\mathcal{E}_p}$ and $\phi_{I_{\mathcal{E}_p}}$ is the finite element shape function associated with node $I_{\mathcal{E}_p}$.
Thanks to the arbitrariness of $\delta u_{i,I_{\mathcal{E}_p}}$ and the fact that by definition $\delta u_{i,I_{\mathcal{E}_p}}=0$ on ${\mathcal{E}_p \cap \partial \Omega_{mpm,disp}}$, the equation for the $I_{\mathcal{E}_p}$-th entry of the local residual vector associated with MPM contributions is obtained from Eq. \eqref{discreteWeakParticles} 
\begin{equation}
\begin{aligned}\label{residualFormLocal}
 r_{mpm,i,I_{\mathcal{E}_p}} &\equiv \sum_{p=1}^{N_p} m_p \Big[ \ddot{u}_{i,p} \phi_{I_{\mathcal{E}_p}}(\bm{x}_p)  
+ \Big( \sum_{j=1}^{d} \sigma^{s}_{ij,p} \dfrac{\partial \phi_{I_{\mathcal{E}_p}}}{\partial x_j}(\bm{x}_p)  \Big)\\
&- b_{i,p} 
%+ \widetilde{t}_{i,p}^{s,k+1} h^{-1} 
\phi_{I_{\mathcal{E}_p}}(\bm{x}_p) \Big]  =0, \quad I_{\mathcal{E}_p} = 1, \ldots N_{\mathcal{E}_p}.
\end{aligned}
\end{equation}
In general, it may not be necessary to interpolate from the grid nodes to the particles to evaluate the body force at the particles. For instance, if the body force is a gravitational force, then no interpolation is necessary and its value at the particles can be computed directly. Therefore, the values of the body force remain evaluated at the particles, and
interpolation may be performed if needed.
The term $\ddot{u}_{i,p}$ in Eq. \eqref{discreteWeakParticles} is computed with the Newmark scheme in Eq. \eqref{Newmark3}.
Specifically, $\ddot{u}_{i,p}$ is given by
\begin{equation}
\begin{aligned}\label{Newmark4}
 \ddot{u}_{i,p} &= \dfrac{1}{\beta \, \Delta t^2} \, \widetilde{u}_{i,g}(\bm{x}_p)-\dfrac{1}{\beta \Delta t} \, \dot{u}^{n}_{i,p} - \dfrac{1-2\beta}{2 \beta} \, \ddot{u}^{n}_{i,p},
\end{aligned}
\end{equation}
where $\widetilde u_{i,g}(\bm{x}_p)$ denotes the reference background grid displacement value at $\bm{x}_p$, defined as
\begin{align}\label{dispGrid}
%  u_{i,g}(\bm{x}_p) \equiv \sum_{I_{\mathcal{E}_p}=1}^{N_{\mathcal{E}_p}} u_{i,I_{\mathcal{E}_p}} \phi_{I_{\mathcal{E}_p}}(\bm{x}_p).
  \widetilde{u}_{i,g}(\bm{x}_p) \equiv \sum_{I_{\mathcal{E}_p}=1}^{N_{\mathcal{E}_p}} \widetilde{u}_{i,I_{\mathcal{E}_p}} \widetilde{\phi}_{I_{\mathcal{E}_p}}(\bm{x}_p),
 \end{align}
 where $\widetilde{u}_{i,I_{\mathcal{E}_p}}$ is as in Eq. \eqref{nodalUTilde}.
The value of $\sigma^{s}_{ij,p}$ in Eq. \eqref{residualFormLocal} refers to the specific Cauchy stress obtained with a partial interpolation from the grid nodes, as in \cite{guilkey2003implicit}.
Recalling the constitutive law in Eq. \eqref{CauchyNeoHookean}, $\sigma^{s}_{ij,p}$ is given by
\begin{align}\label{Cauchy}
\bm{\sigma}^{s}_p =   \dfrac{\lambda \,\, \log(J_p)}{\rho \,\, J_p} \, \bm{I} + \dfrac{\mu}{J_p} \, \Big( \bm{B}_p - \bm{I}\Big),
\end{align}
where $J_p = \det(\bm{F}_p)$, $\bm{B}_p = \bm{F}_p \, (\bm{F}_p)^T $ and 
\begin{align}\label{defGradTime}
 \bm{F}_p = \Big( \widetilde{\nabla} \widetilde{\bm{u}}_g(\bm{x}_p) + \bm{I} \Big) \, \bm{F}_p^{n}.
\end{align}
In Eq. \eqref{defGradTime}, $\widetilde{\nabla}$ represents the Del operator with respect to the reference configuration and  $\widetilde{\nabla} \widetilde{\bm{u}}_g(\bm{x}_p)$ is the background grid displacement gradient with respect to the reference configuration, evaluated at $\bm{x}_p$, defined as
\begin{align}\label{gradNablaRef}
 \widetilde{\nabla} \widetilde{u}_{ij,g}(\bm{x}_p) \equiv  \sum_{I_{\mathcal{E}_p}=1}^{N_{\mathcal{E}_p}} \widetilde{u}_{i,I_{\mathcal{E}_p}} \dfrac{\partial \widetilde{\phi}_{I_{\mathcal{E}_p}}}{\partial \widetilde{x}_j}(\bm{x}_p).
\end{align}

It is important to stress that if no coupling is considered, the undeformed configuration coincides with the reference configuration, since the reset displacement field is identically zero.
This can be easily recalled considering Eq. \eqref{relationReset}.

The next step consists of determining the scaling factors to be used in the assembly of the {\it{soft stiffness}} matrix. 
Such factors depend on flags $M_{\mathcal{E}}$ assigned to each element of the active background grid $\mathcal{T}_{ab}$. The value of a flag depends on the element's relative position with respect to the MPM particles. 
Let $N_{\mathcal{E}}$ be the total number of degrees of freedom associated with an element $\mathcal{E}$ on the active background grid.
For instance, for bi-quadratic quadrilateral elements $N_{\mathcal{E}} = 9$. \giachi{See for instance Appendix A in \cite{capodaglio2017particle} for a sketch of the Lagrangian bi-quadratic triangle and quadrilateral with associated degrees of freedom.}
For simplicity, we only consider the bi-quadratic case, but the following reasoning can be easily extended to linear and serendipity elements, as well as to higher dimensions.
The value of the flag $M_{\mathcal{E}}$ is initialized to $N_{\mathcal{E}}$ for all elements in the active background grid.
Let $\mathcal{E}$ be any of such elements. Then the value of $M_{\mathcal{E}}$ is modified according to the following criterion: for any node $I_{\mathcal{E}}$ of $\mathcal{E}$ that also belongs to an element of $\mathcal{T}_{ib}$, 
$M_{\mathcal{E}}$ is decreased by a unit.
Hence, for instance, bi-quadratic quadrilateral elements that are surrounded by elements that do not enclose particles will have $M_{\mathcal{E}}=1$.\\

Next, a loop on all elements that are part of the active background grid is carried out in the standard finite element fashion, to compute the soft stiffness contribution of the local residual 
\begin{equation}
\begin{aligned}\label{softResidualLocal}
 {r}_{soft,i,I_{\mathcal{E}}} \equiv  \mu  \, C_{\mathcal{E}} \int_{\mathcal{E}}  
 \Big[\widetilde{\nabla} \bm{\widetilde{u}} + \big(\widetilde{\nabla} \bm{\widetilde{u}} \big)^T \Big]_i \cdot \widetilde{\nabla} \widetilde{\phi}_{I_{\mathcal{E}}} dV,
 \, \mbox{for} \quad I_{\mathcal{E}} = 1, \ldots, N_{\mathcal{E}}.
\end{aligned}
\end{equation}
The integral in Eq. \eqref{softResidualLocal} is approximated by means of a quadrature rule on the element $\mathcal{E}$, where Gauss points are used as quadrature points.
\giachi{The value of $C_{\mathcal{E}}$ is a measure of the magnitude of the soft stiffness contribution and it depends on the flags $M_{\mathcal{E}}$ previously determined. This is easily explained by the fact that the larger $M_{\mathcal{E}}$, the more elements filled with particles surround $\mathcal{E}$, and the smaller soft stiffness contribution is needed to stabilize the equations within $\mathcal{E}$.
As a general rule, $C_{\mathcal{E}}$ is a non-increasing function of $M_{\mathcal{E}}$, and it is zero for any element fully surrounded by elements containing particles. 
The soft stiffness improves stability while decreasing accuracy. Accuracy decreases because a contribution that is not associated with the MPM problem is added to the system. As mentioned in the introduction, it is necessary to strike a balance between the need for accuracy and stability.  A value of $C_{\mathcal{E}} = 0$ is appropriate for elements that are fully surrounded by elements with particles but it may not be sufficient for those elements on the boundary of the MPM body. This is because they may not be filled enough with particles.  
It is recommended to keep $C_{\mathcal{E}}$ as small as possible, for instance in the interval $[10^{-2}, 10^{-8}].$
In the numerical simulations, we do not change the mass of the particles, but we test our model for increasing values of the Young's modulus $E$.
As a consequence, the stiffness of the body increases, since $C_{\mathcal{E}}$ is multiplied by $\mu$, which is proportional to $E$, while the gravitational force acting on the body remains the same.
If the constant $C_{\mathcal{E}}$ remains the same, the artificial stiffness proportional to $\mu\, C_{\mathcal{E}}$ would increase. Hence, it is reasonable to decrease the value of $C_{\mathcal{E}}$ at least proportionally to the increase of the Young's modulus. A dimensionless analysis of the equation would confirm this straightforwardly.
Examples of values for $C_{\mathcal{E}}$ will be given in Section \ref{numRes}.}

 Note that Eq. \eqref{residualFormLocal} and Eq. \eqref{softResidualLocal} only involve nodes that belong to the active background grid $\mathcal{T}_{ab}$, while the equation system at this stage 
 is actually defined and solved for all the nodes of $\mathcal{T}_b$. 
 In such a way, there is no need to construct a new background grid at each instant of time, 
 rather part of the background grid becomes active depending on the position of $\Omega_{mpm}$, as already explained above.
 For all nodes that belong to $\mathcal{I}_{ib}$ (so exclusively to the inactive background grid),  a fictitious equation is assembled using a lumped mass matrix, as it is shown next.\\

\item {\bf{Assembly of the inactive background grid contributions}}.
\newline
Here, a loop on all the elements of the inactive background grid is carried out in a standard finite element fashion.
\giachi{Let $\mathcal{E}$ be any element in $ \mathcal{T}_{ib}$. Then, for any node $I_\mathcal{E} \in \mathcal{I}_{ib}$, the local residual is computed as the product of a lumped mass matrix with the nodal displacement vector, as follows}
\begin{equation}
\begin{aligned}\label{massResidualLocal}
 {r}_{mass,i,I_{\mathcal{E}}} &\equiv \int_{\widehat{\mathcal{E}}} u_{i,I_{\mathcal{E}}}\widehat{\phi}_{I_{\mathcal{E}}} dV   \approx u_{i,I_{\mathcal{E}}} \Big(\sum_{i_g=1}^{N_{\mathcal{E},g}} \widehat{\omega}_{i_g}  \widehat{\phi}_{I_{\mathcal{E}}}(\bm{x}_{i_g})\Big), 
\end{aligned}
\end{equation}
\giachi{for $I_{\mathcal{E}} = 1, \ldots, N_{\mathcal{E}}$, $I_{\mathcal{E}} \in \mathcal{T}_{ib}$.
Recall that $\mathcal{I}_{ib}$ contains only the nodes that belong exclusively to the inactive background grid, so nodes that are shared with elements that include at least one particle are excluded from the inactive background grid contribution.}
In Eq. \eqref{massResidualLocal}, for $i_g = 1, \ldots, {N_{\mathcal{E},g}}$, $\bm{x}_{i_g}$ is a Gauss point of the undeformed element $\widehat{\mathcal{E}}$ associated with  $\mathcal{E}$, $\widehat{\omega}_{i_g}$ is the quadrature weight associated with $\bm{x}_{i_g}$, $u_{i,I_{\mathcal{E}}}$ is $i$-th component of the nodal displacement value at the local node $I_{\mathcal{E}}$, and $\widehat{\phi}_{I_{\mathcal{E}}}$ is the shape function associated with node $I_{\mathcal{E}}$ of the undeformed element $\widehat{\mathcal{E}}$.\\

\item {\bf{Differentiation of the residual vector}}.
\newline
Let $\bm{r}_{mpm, \mathcal{E}_p} = [r_{mpm,1,1},r_{mpm,1,2}, \ldots, r_{mpm,d,N_{\mathcal{E}_p}}]$ denote the local residual vector associated with the MPM contributions from element $\mathcal{E}_p$.
The global MPM residual vector is denoted by 
$$\bm{r}_{mpm} = [r_{mpm,1,1},r_{mpm,1,2}, \ldots, r_{mpm,d,N_g}]$$ and
 is obtained in the standard finite element fashion, adding the entries of the local MPM residual vectors associated with the same global nodes.
 In the same way, the global soft residual vector $\bm{r}_{soft}$ and the global mass residual vector $\bm{r}_{mass}$ are constructed.
 It is important from an implementational point of view to ensure that the three global residual have matching dimensions and that they are all initialized to zero.
 Next, the global residual vector $\bm{r}$ for the background grid system is defined as
 \begin{align}\label{globalResidual}
 \bm{r} =  \bm{r}_{mass} + \bm{r}_{soft} + \bm{r}_{mpm}.
 \end{align}
Let $\bm{u}^{k+1}=[u^{k+1}_{1,1}, u^{k+1}_{1,2}, \ldots, u^{k+1}_{d,N_g}]$ represent the vector of displacement values at the grid nodes at iteration $k+1$, then $\bm{r}(\bm{u}^{k+1})$ is a nonlinear function of $\bm{u}^{k+1}$ and
to determine the values of the displacement at the grid nodes, it is necessary to solve the nonlinear equation
\begin{align} \label{nonlinearEq}
 \bm{r}(\bm{u}^{k+1}) = 0.
\end{align}
Writing $\bm{u}^{k+1} = \bm{u}^{k} + \Delta \bm{u}^{k+1}$ and expanding Eq.\eqref{nonlinearEq} in a Taylor series around $\bm{u}^{k}$, the following equation is obtained 
\begin{align}
 0 = \bm{r}(\bm{u}^{k+1}) = \bm{r}(\bm{u}^{k}) + \dfrac{\partial \bm{r}(\bm{u}^{k})}{\partial \bm{u}} \Delta \bm{u}^{k+1} + O(({\Delta \bm{u}^{k+1}})^2).
\end{align}
Neglecting the higher order terms, a linear system of equations is recovered
\begin{align}\label{linSys}
 \bm{K}(\bm{u}^k) \Delta \bm{u}^{k+1} = - \bm{r}(\bm{u}^{k}),
\end{align}
where $\bm{K}(\bm{u}^k) \equiv \dfrac{\partial \bm{r}(\bm{u}^{k})}{\partial \bm{u}}$ is the background grid stiffness matrix,  $\Delta \bm{u}^{k+1}$ is the solution increment vector and the forcing term $- \bm{r}(\bm{u}^{k})$ is the negative residual vector associated with the solution at the previous iteration.
In the numerical implementation, the matrix $\bm{K}(\bm{u}^k)$ is obtained using automatic differentiation provided by the Adept library \cite{hogan2014fast}.
Hence, once the residual has been assembled, no explicit computation has to be carried out to obtain such a matrix.\\
After an appropriate reordering of the grid nodes, Eq. \eqref{linSys} can be written as the following block system
\begin{align} \label{blockSys}
 \begin{bmatrix}
   \bm{K}_{ab}(\bm{u}_{ab}^k)      \quad &    \bm{0}  \\
    \\
   \bm{0}      \quad &     \bm{M}_{ib}  \\
\end{bmatrix}
 \begin{bmatrix}
   \Delta \bm{u}_{ab}^{k+1}       \\
    \\
    \Delta \bm{u}_{ib}^{k+1}       
\end{bmatrix} 
=
 \begin{bmatrix}
    - \bm{r}_{ab}(\bm{u}_{ab}^{k})       \\
    \\
    - \bm{r}_{ib}(\bm{u}_{ib}^{k})     
\end{bmatrix},
\end{align}
where
\begin{equation}
\begin{aligned}
&\bm{K}_{ab}(\bm{u}_{ab}^k)  = \bm{K}_{soft}(\bm{u}_{ab}^k) + \bm{K}_{mpm}(\bm{u}_{ab}^k) ,\\
& \bm{r}_{ab}(\bm{u}_{ab}^{k}) = \bm{r}_{soft}(\bm{u}_{ab}^{k}) + \bm{r}_{mpm}(\bm{u}_{ab}^{k}),\\
&\bm{K}_{soft}(\bm{u}_{ab}^k) \equiv \dfrac{\partial \bm{r}_{soft}(\bm{u}_{ab}^{k})}{\partial \bm{u}_{ab}}, \quad
\bm{K}_{mpm}(\bm{u}_{ab}^k) \equiv \dfrac{\partial \bm{r}_{mpm}(\bm{u}_{ab}^{k})}{\partial \bm{u}_{ab}},\\
& \bm{r}_{ib}(\bm{u}_{ib}^{k}) = \bm{r}_{mass}(\bm{u}_{ib}^{k}), \quad \bm{M}_{ib} = \dfrac{\partial \bm{r}_{mass}(\bm{u}_{ib}^{k})}{\partial \bm{u}_{ib}}.
\end{aligned}
\end{equation}

\item {\bf{Solution of the background grid system}}.
\newline
Let $\Delta \bm{u}^{k+1}$ denote the solution of system \eqref{blockSys}.
With such an information, the displacement at the grid is advanced 
$\bm{u}^{k+1} = \bm{u}^{k} + \Delta \bm{u}^{k+1}$. If the displacement falls below a prescribed tolerance, the algorithm moves to the next stage, otherwise further iterations are carried out.\\

\end{enumerate}
\item {\bf{Update of Particle Instances}}.
\newline
In the last stage of the algorithm, the position, velocity, acceleration and deformation gradient of the particles are updated.
As for the assembly procedure, this step begins with a loop on the particles. Then, for a given particle $p$ in the loop, the element $\mathcal{E}_p$ that hosts $p$ is again extracted.
The deformation gradient is updated in the following way
\begin{align}\label{updateFp}
\bm{F}_p^{n+1} = ( \widetilde{\nabla} \bm{\widetilde{u}}^{n+1}_g(\bm{x}_p) + \bm{I}) \, \bm{F}_p^{n}, 
\end{align}
where
\begin{align}\label{dispGrid2}
%  u_{i,g}(\bm{x}_p) \equiv \sum_{I_{\mathcal{E}_p}=1}^{N_{\mathcal{E}_p}} u_{i,I_{\mathcal{E}_p}} \phi_{I_{\mathcal{E}_p}}(\bm{x}_p).
  \bm{\widetilde{u}}_{g}^{n+1}(\bm{x}_p) \equiv \sum_{I_{\mathcal{E}_p}=1}^{N_{\mathcal{E}_p}} \bm{\widetilde{u}}^{n+1}_{I_{\mathcal{E}_p}} \widetilde{\phi}_{I_{\mathcal{E}_p}}(\bm{x}_p).
 \end{align}
The particle displacement is obtained via interpolation from the displacement $\widetilde{\bm{u}}_I$ at the nodes of $\mathcal{E}_p$. Namely, $\bm{u}^{n+1}_p \equiv \bm{\widetilde{u}}_{g}^{n+1}(\bm{x}_p)$.
With this information and the particle instances at the previous instant of time, velocity and acceleration are updated using the Newmark scheme in Eq. \eqref{Newmark1} and Eq. \eqref{Newmark3} respectively.
Finally, the particle positions are updated as $\bm{x}^{n+1}_p = \bm{x}^{n}_p + \bm{u}^{n+1}_p$.
\end{itemize}

\section{Monolithic coupling between MPM and FEM}\label{couplingMPMFEM}
In this section, the solid-solid monolithic coupling between the proposed implicit material point method and the finite element method is described. The two solid bodies are treated as a single continuum, and a shared finite element grid is used for the FEM body and the MPM background grid. In this way, the MPM background grid follows the solid deformations 
and the interface between MPM and FEM bodies is automatically tracked. 
With this monolithic approach, there is no need for time consuming contact algorithms because the MPM particles 
do not penetrate the FEM body (conservation of mass). In addition, the continuity of the normal stress on the shared interface
is automatically satisfied (conservation of momentum).
The monolithic coupling used in this work is highly inspired by monolithic coupling strategies for fluid-structure interaction (FSI) 
problems \cite{calandrini2018valve, aulisa2018monolithic, calandrini2017magnetic, aulisa2017fluid, aulisa2017fluid2}.
As a matter of fact, the MPM background grid is used in the same way as the finite element grid employed to discretize the fluid domain in FSI problems. In such problems, the displacement of the fluid grid at the interface is required to match the displacement of the solid grid. The same requirement is enforced here between the MPM background grid and the FEM solid grid. 
However, there is a major difference between the approach adopted in this work and the one used to model FSI problems.
Indeed, for any node of the background grid that does not belong to the FEM body, the displacement is reset to zero after every time step, as the MPM algorithm requires. On the other hand, such a reset does not occur for the fluid displacement in FSI problems. Another major difference is that of course no actual fluid equation is solved on the background grid in the present case, and the active background grid is only used for interpolation during the solution of the MPM equations.

The weak system of equations solved in the coupled case is the following

\begin{equation}
\begin{aligned} \label{coupledMPM}
&\int_{\Omega_{ab}} \rho_{mpm} \, \Big(\ddot{\bm u}_{mpm} \cdot \delta \bm u_{mpm} \, + \bm \sigma_{mpm}^{s} : \nabla \delta \bm u_{mpm} \, -  \, \bm b_{mpm} \cdot \delta  \bm u_{mpm} \Big) \, dV \\
&\quad - \int_{\Gamma_a} \rho_{mpm} (\bm\sigma_{mpm}^{s} \cdot \delta \bm u_{mpm} ) \cdot \bm n_{mpm} \ = 0, \qquad \forall \delta \bm u_{mpm} \in \Delta U_{mpm},   \\
&\int_{\Omega_{fem}}  \rho_{fem} \ddot{\bm u}_{fem} \cdot \delta \bm u_{fem} \, + \bm \sigma_{fem} : \nabla \delta \bm u_{fem} \, -  \, \rho_{fem} \bm b_{fem} \cdot \delta  \bm u_{fem}  \, dV \\
& - \int_{\Gamma_a} (\bm\sigma_{fem} \cdot \delta \bm u_{fem} ) \cdot \bm n_{fem} \ = 0, \qquad \forall \delta \bm u_{fem} \in \Delta U_{fem}, 
\end{aligned}
\end{equation}
where $\Delta U_{fem}$ represents the set of the test functions for the FEM weak formulation.
Since two elastic bodies are simulated, the momentum equation is the same for both the MPM and the FEM body and the Cauchy stress is still given by Eq. \eqref{CauchyNeoHookean} for both solids.
Recall from Section \ref{MPMFEMdomains} that $ \Gamma_a =  \partial \Omega_{ab} \cap \partial \Omega_{fem} $ denotes the interface between the MPM and the FEM domains.
The continuity of mass and momentum between the equations in system \eqref{coupledMPM} are satisfied. Namely,
\begin{equation}
\begin{aligned}
 & \bm{u}_{mpm} = \bm{u}_{fem} \\
 & \left. (\bm{\sigma}\cdot \bm{n}) \right|_{fem,\Gamma_a} + \left. (\bm{\sigma}\cdot \bm{n}) \right|_{mpm, \Gamma_a} = 0,\\
\end{aligned}
 \end{equation}
where the second equation means that the FEM and the MPM values of $(\bm{\sigma}\cdot \bm{n})$ match on $\Gamma_a$. 
Due to the monolithic approach, the equations for the MPM body and for the FEM body are solved at the same time with a unique assembly for the two.
The assembly procedure is carried out exactly in the same way as in Section \ref{numAlg} for the nodes on the MPM background grid $\mathcal{T}_b$. 
\giachi{If $\mathcal{T}_{fem}$ denotes the finite element grid used to discretize $\Omega_{fem}$, let $\mathcal{I}_{fem}$  be the set of grid points of $\mathcal{T}_{fem}$. The inactive grid contributions are given only to the nodes in $ \mathcal{I} \setminus (\mathcal{I}_{ab} \cup \mathcal{I}_{fem})$, so no coupling between inactive nodes and FEM nodes exists.}
For any node in $\mathcal{I}_{fem}$, the assembly procedure is carried out in a standard finite element fashion, using Gauss points as quadrature points.
Recalling the constitutive law in Eq. \eqref{CauchyNeoHookean}, for a given Gauss point $i_g$, we define $\sigma^{s}_{ij,{i_g}}$ by
\begin{align}\label{CauchyGauss}
\bm{\sigma}^{s}_{i_g} \equiv  \dfrac{\lambda_{fem} \,\, \log(J_{i_g})}{\rho_{fem} \,\, J_{i_g}} \, \bm{I} + \dfrac{\mu_{fem}}{J_{i_g}} \, \Big( \bm{B}_{i_g} - \bm{I}\Big),
\end{align}
where $J_{i_g} = \det(\bm{F}_{i_g})$, $\bm{B}_{i_g} = \bm{F}_{i_g}\, (\bm{F}_{i_g})^T $ and 
\begin{align}\label{defGradTimeGauss}
 \bm{F}_{i_g} = \Big( \widehat{\nabla} \bm{u}_g(\bm{x}_{i_g}) + \bm{I} \Big).
\end{align}
In Eq. \eqref{defGradTimeGauss}, $\widehat{\nabla} \bm{u}_g(\bm{x}_{i_g})$ represents the background grid displacement gradient with respect to the undeformed finite element configuration, evaluated at $\bm{x}_{i_g}$, given by
\begin{align}\label{gradNablaRefGauss}
 \widehat{\nabla} u_{ij,g}(\bm{x}_{i_g}) = \sum_{I_{\mathcal{E}}=1}^{N_{\mathcal{E}_g}} u_{i,I_{\mathcal{E}}} \dfrac{\partial \widehat{\phi}_{I_{\mathcal{E}}}}{\partial \widehat{x}_j}(\bm{x}_{i_g}).
\end{align}
For any  $\mathcal{E} \in \mathcal{T}_{fem}$, the local residual is computed as  
\begin{equation}
\begin{aligned}\label{femResidualLocal}
 {r}_{fem,i,I_{\mathcal{E}}} &\equiv \sum_{i_g=1}^{N_{{\mathcal{E}}_g}} \omega_{i_g} \Big[ \ddot{u}_{i,i_g} \phi_{I_{\mathcal{E}}}(\bm{x}_{i_g})  
+ \Big( \sum_{j=1}^{d} \sigma^{s}_{ij,i_g} \dfrac{\partial \phi_{I_{\mathcal{E}}}}{\partial x_j}(\bm{x}_{i_g})  \Big)\\
&- \dfrac{\rho_{fem} \,  b_{i,i_g}}{J_{i_g}} 
%+ \widetilde{t}_{i,p}^{s,k+1} h^{-1} 
\phi_{I_{\mathcal{E}}}(\bm{x}_{i_g}) \Big]  =0, \,\, \mbox{for} \,\, I_{\mathcal{E}}=1, \ldots, N_{\mathcal{E}}.
\end{aligned}
\end{equation}
Since particles are substituted by Gauss points in the FEM assembly, the Newmark scheme in Eq. \eqref{Newmark3} for the FEM body becomes
\begin{align}\label{NewmarkGauss}
 \ddot{\bm{u}}^{n+1}_{i_g} = \dfrac{1}{\beta \, \Delta t^2} \, \Big({\bm{u}}^{n+1}_{i_g} - {\bm{u}}^{n}_{i_g}\Big)
 -\dfrac{1}{\beta \Delta t} \,\, \dot{\bm{u}}^{n}_{i_g} - \dfrac{1-2\beta}{2 \beta}\ddot{\bm{u}}^{n}_{i_g},
\end{align}
where 
\begin{align}\label{dispGridGauss}
 {\bm{u}}^{n+1}_{i_g} = \bm{u}^{n+1}(\bm{x}_{i_g}) \equiv \sum_{I_{\mathcal{E}}=1}^{N_{\mathcal{E},g}} \bm{u}^{n+1}_{I_{\mathcal{E}}} \phi_{I_{\mathcal{E}}}(\bm{x}_{i_g}).
 \end{align}
 The displacement, velocity and acceleration at the Gauss point at time $t^n$ in Eq. \eqref{NewmarkGauss} are computed in an analogous way
 as the displacement at the Gauss point at time $t^{n+1}$ in Eq. \eqref{dispGridGauss}.
 The coupled monolithic block system is again block diagonal and is given by
\begin{align} \label{coupledSysMatrix}
 \begin{bmatrix}
   \bm{K}_{ab}(\bm{u}^k)  & \bm{K}_{ab,fem}(\bm{u}^k)  &  \bm{0}  \\ \\
   \bm{K}_{fem,ab}(\bm{u}^k)  &  \bm{K}_{fem}(\bm{u}^k)  & \bm{0}  \\ \\
   \bm{0}   &  \bm{0} & \bm{M}_{ib}  \\
\end{bmatrix}
 \begin{bmatrix}
   \Delta \bm{u}_{ab}^{k+1}       \\ \\
    \Delta \bm{u}_{fem}^{k+1} \\ \\
    \Delta \bm{u}_{ib}^{k+1}       
\end{bmatrix} 
=
 \begin{bmatrix}
    - \bm{r}_{ab}(\bm{u}^{k})       \\ \\
    - \bm{r}_{fem}(\bm{u}^{k}) \\ \\
    - \bm{r}_{ib}(\bm{u}^{k})     
\end{bmatrix},
\end{align}
 where
 \giachi{
\begin{equation}
\begin{aligned}
&\bm{K}_{ab}(\bm{u}^k)  = \bm{K}_{soft}(\bm{u}^k) + \bm{K}_{mpm}(\bm{u}^k) ,\\
& \bm{r}_{ab}(\bm{u}^{k}) = \bm{r}_{soft}(\bm{u}^{k}) + \bm{r}_{mpm}(\bm{u}^{k}),\\
&\bm{K}_{soft}(\bm{u}^k) \equiv \dfrac{\partial \bm{r}_{soft}(\bm{u}^{k})}{\partial \bm{u}_{ab}}, \quad
\bm{K}_{mpm}(\bm{u}^k) \equiv \dfrac{\partial \bm{r}_{mpm}(\bm{u}^{k})}{\partial \bm{u}_{ab}},\\
&\bm{K}_{ab,fem}(\bm{u}^k) \equiv \dfrac{\partial \bm{r}_{ab}(\bm{u}^k)}{\partial \bm{u}_{fem}}, \quad
\bm{K}_{fem,ab}(\bm{u}^k) \equiv \dfrac{\partial \bm{r}_{fem}(\bm{u}^k)}{\partial \bm{u}_{ab}},\\
& \bm{r}_{ib}(\bm{u}^{k}) = \bm{r}_{mass}(\bm{u}^{k}), \quad \bm{M}_{ib} = \dfrac{\partial \bm{r}_{mass}(\bm{u}^{k})}{\partial \bm{u}_{ib}}.
\end{aligned}
\end{equation}
}
The (1,2) and (2,1) blocks represent the coupling between the active background grid and the FEM body nodes.
 After the solution of the monolithic system, the nodal values of velocity and acceleration at the grid nodes in $\mathcal{I}_{fem}$ are updated using the Newmark scheme
 \begin{equation}
  \begin{aligned}
   	\ddot{\bm{u}}_{I}^{n+1} &= \dfrac{1}{\beta \Delta t^2} \giachi{\Big({\bm{u}}_{I}^{n+1} - {\bm{u}}_{I}^{n} \Big)} - \dfrac{1}{\beta \Delta t}\dot{\bm{u}}_{I}^{n} - \dfrac{1 - 2 \, \beta}{2  \beta} \ddot{\bm{u}}_{I}^{n}, \\
	\dot{\bm{u}}_{I}^{n+1} &= \dot{\bm{u}}_{I}^{n} + \Delta t  \Big( (1 - \gamma)  \ddot{\bm{u}}_{I}^{n} + \gamma \, \ddot{\bm{u}}_{I}^{n+1}\Big).
  \end{aligned}
 \end{equation}

\section{Receding Contact Force Correction}\label{receding}
\giachi{The MPM domain may be subject to roto-translation and large deformations, and it is possible for its boundary 
to come into contact with the domain boundary where a Dirichlet boundary condition is imposed 
or with the FEM solid boundary.
These situations require the contact force modeling to be carefully described. 
The contact force between two surfaces acts only when they touch, and a mutual pressure is applied.
During the receding phase, when the bodies detach, the contact force must be zero. Moreover, compenetration between the touching domains should never occur.
The proposed monolithic formulation never allows for compenetration, and the force balance at the interface is automatically satisfied in the touching phase of the contact. However, in the receding phase, the contact suffers of small to moderate {\emph{sticky}} effects, that grow with increasing values of the Young's modulus of the MPM body.}

\giachi{To avoid the artificial stickiness in the receding phase, a simple correction to the contact description is embedded in the implicit monolithic formulation.
The correction requires a small modification in the MPM assembly for those elements of the active background grid for which a face is shared with an element of the FEM body, or Dirichlet boundary conditions are imposed on it.
The nodal displacement values of such elements receive an additional contribution for each enclosed particle that moves away from the face.
The idea is to reduce the reaction force exerted by the face in the evaluation of $\bm{F}_p$ in Eq. \eqref{defGradTime}. 
In particular, a reduction of the contribution of $\widetilde{\nabla} \widetilde{\bm{u}}_g(\bm{x}_p)$ to the deformation gradient   $\bm{F}_p$ is obtained.}
%%%%%%%%%%%%%%%%%%%%%%%%%%%%%%%%%%%%%%%%%%%%%%%%%%%%%%%%%%%%%%%%%%%%%%%%%%%%%%%%%%%%%%%%%%%%%%%%%%%%%%%%%%%%%%%%%%%%%%%%%%%%%%%
\begin{figure}[t]
\centering
\includegraphics[scale=0.8]{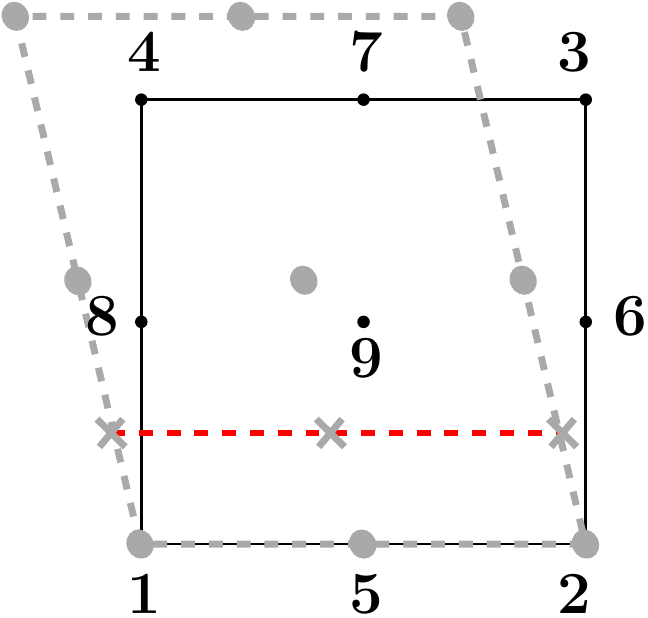}
\caption{\giachi{Schematics for the receding contact force algorithm}}\label{deformed}
\end{figure}
%%%%%%%%%%%%%%%%%%%%%%%%%%%%%%%%%%%%%%%%%%%%%%%%%%%%%%%%%%%%%%%%%%%%%%%%%%%%%%%%%%%%%%%%%%%%%%%%%%%%%%%%%%%%%%%%%%%%%%%%%%%%%%%
\giachi{The algorithm has been implemented here only for bi-quadratic Lagrangian quadrilateral elements but it readily extends to other shapes and families of finite elements.
Let $p$ be a particle and let $\mathcal{E}_p$ be the element that encloses it, depicted in Figure \ref{deformed}.
For simplicity, we consider the case where Dirichlet zero boundary conditions have been imposed on the lower face of $\mathcal{E}_p$. Hence, the lower face will remain undeformed in the time interval $(t^n, t^{n+1}]$.
This implies that $\widetilde{\bm{u}}_{1} = \widetilde{\bm{u}}_{2} = \widetilde{\bm{u}}_{5} = \bm 0 $, whereas
the other nodes are free to move, see the dashed gray contour in Figure \ref{deformed}. 
If the particle moves away from the Dirichlet boundary, the values of $\widetilde{\bm{u}}_{I_{\mathcal{E}_p}}$, with ${I_{\mathcal{E}_p}} = 1, \ldots, N_{\mathcal{E}_p}=9$, are modified as follows
\begin{equation}\label{contactCorr}
\begin{aligned}
  & \widetilde{\bm{u}}_{1}^* = (1-C_c) \widetilde{\bm{u}}_{1} + C_c \left(\frac{4}{3} \widetilde{\bm{u}}_{4} - \frac{1}{3} \widetilde{\bm{u}}_{8} \right),\\
  & \widetilde{\bm{u}}_{2}^* = (1-C_c) \widetilde{\bm{u}}_{2} + C_c \left(\frac{4}{3} \widetilde{\bm{u}}_{3} - \frac{1}{3} \widetilde{\bm{u}}_{6} \right),\\
  & \widetilde{\bm{u}}_{5}^* = (1-C_c) \widetilde{\bm{u}}_{5} + C_c \left(\frac{4}{3} \widetilde{\bm{u}}_{7} - \frac{1}{3} \widetilde{\bm{u}}_{9} \right),\\
  & \widetilde{\bm{u}}_{I_{\mathcal{E}_p}}^* =  \widetilde{\bm{u}}_{I_{\mathcal{E}_p}} \text{ for } I_{\mathcal{E}_p} = 3,4,6,7,8,9\,,
\end{aligned}
\end{equation}
with $C_c\in[0,1]$. Note that the terms multiplied by $C_c$ in the above equations represent the correction terms. The new deformation field allows the nodes on the lower face to move, see the dashed red line in Figure \ref{deformed}.
For $C_c=0$, the Dirichlet boundary condition is restored. For $C_c = 1$, the new displacement field is such that 
$\widetilde{\nabla} \widetilde{\bm{u}}_g^*(\bm{x_1}) = \widetilde{\nabla} \widetilde{\bm{u}}_g^*(\bm{x_2}) = \widetilde{\nabla} \widetilde{\bm{u}}_g^*(\bm{x_5}) = \bm 0$, and consequently
for each particle $p$ located on the lower face it would be 
\begin{align}\label{NewdefGradTime}
\bm{F}_p = \Big( \widetilde{\nabla} \widetilde{\bm{u}}^*_g(\bm{x}_p) + \bm{I} \Big) \, \bm{F}_p^{n} = \bm{F}_p^{n},
\end{align}
with no new contribution to the stress tensor. For $C_c \in (0,1)$, the resulting boundary condition can be considered of mixed type. 
The equations in \eqref{contactCorr} can be used also for the case of non zero Dirichlet boundary condition 
or contact of the MPM body with the FEM body. }

\giachi{The value of the constant $C_c$ has been investigated numerically. It is shown in Section \ref{numRes} that if an MPM body rolls on a boundary (or on a FEM body), one can choose $C_c = 0.5$ regardless of the material type, the grid resolution and the coefficients $C_{\mathcal{E}}$ for the soft stiffness matrix. For the case of a bouncing MPM object on a FEM plate, the numerical results show sensitivity to $C_c$, and a different value has to be chosen any time one of the parameters changes. However, it will be shown that the range of values of $C_c$ remains small.}
 
 \section{Numerical Results}\label{numRes}
 Numerical results are presented in this section to showcase the performances of the algorithms here described.
 First, the implicit MPM scheme is tested, followed by the monolithic coupling between MPM and FEM.
 When possible, the numerical results are compared to analytic solutions to show the great accuracy of the methods presented.
 For simplicity, only 2-dimensional numerical tests are performed, even if the methods developed in this work readily apply to 3-dimensional problems as well.
 
 \subsection{Initialization of the MPM Bodies}
 \giachi{In the following numerical examples, two different types of MPM bodies are considered, a disk and a beam. Both bodies have a uniform particle distribution in their core
 and a progressively refined distribution moving from an a priori chosen interior surface towards the boundary.
 Recall that the particle mass is defined as $m_p = \rho_0 V_p$, where $\rho_0$ is the initial density of the MPM body and
 $V_p$ is the particle volume. This volume has a value that depends on the location of the particle within the particle distribution used to discretized the body.
 Two different initialization procedures are carried out depending on the specific body.}
 
 \subsubsection{Initialization of the disk}
 \giachi{The input parameters for the disk are: 
  the coordinates of the center of the disk $(x_c,y_c)$, 
 the radius of the disk $R$ , 
 the radius of the interior disk in which the uniform distribution is built $R_0$ , 
 and $N_{\theta_0}$, which is the number of particles located within the disk of radius $R_0$.
 The number of boundary layers $N_b$ between the circles can be obtained after $R$, $R_0$, and $N_{\theta_0}$ have been chosen. For each particle, the output parameters are its initial coordinates 
 and its volume $V_p$.}
  
\subsubsection{Initialization of the beam}
 \giachi{The input parameters for the beam are: 
 the coordinates of the center point of the left boundary $(x_0,y_0)$, 
  the height $H$ and length $L$ of the beam, 
 the height $H_0$  of the interior beam in which a uniform distribution is built, 
 and the number $N_H$  of particle rows in the interior beam.
 The number of boundary layers $N_b$ between the beams can be obtained once $H$, $H_0$ and $N_H$ are chosen.
 For each particle, the output parameters are again its initial coordinates 
 and its volume $V_p$.}
 
\giachi{The pseudocode for the initialization of the two MPM bodies is given in Appendix A.
We remark that the performances of the method do not depend on the specific particle distribution adopted in this work and other distributions may be possible.}
 
 \subsection{Tests for the implicit MPM}\label{numTestsMPM}
 
\begin{figure}[htb]
\centering
\includegraphics[scale=0.9]{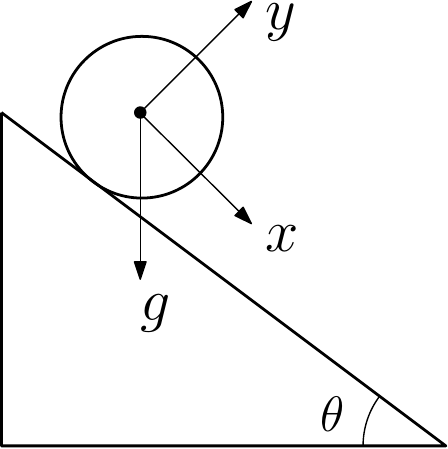}
\caption{Schematics of the rolling disk test.}\label{rolling}
\end{figure}

\begin{figure}[htb]
\centering
I) \includegraphics[scale=0.55]{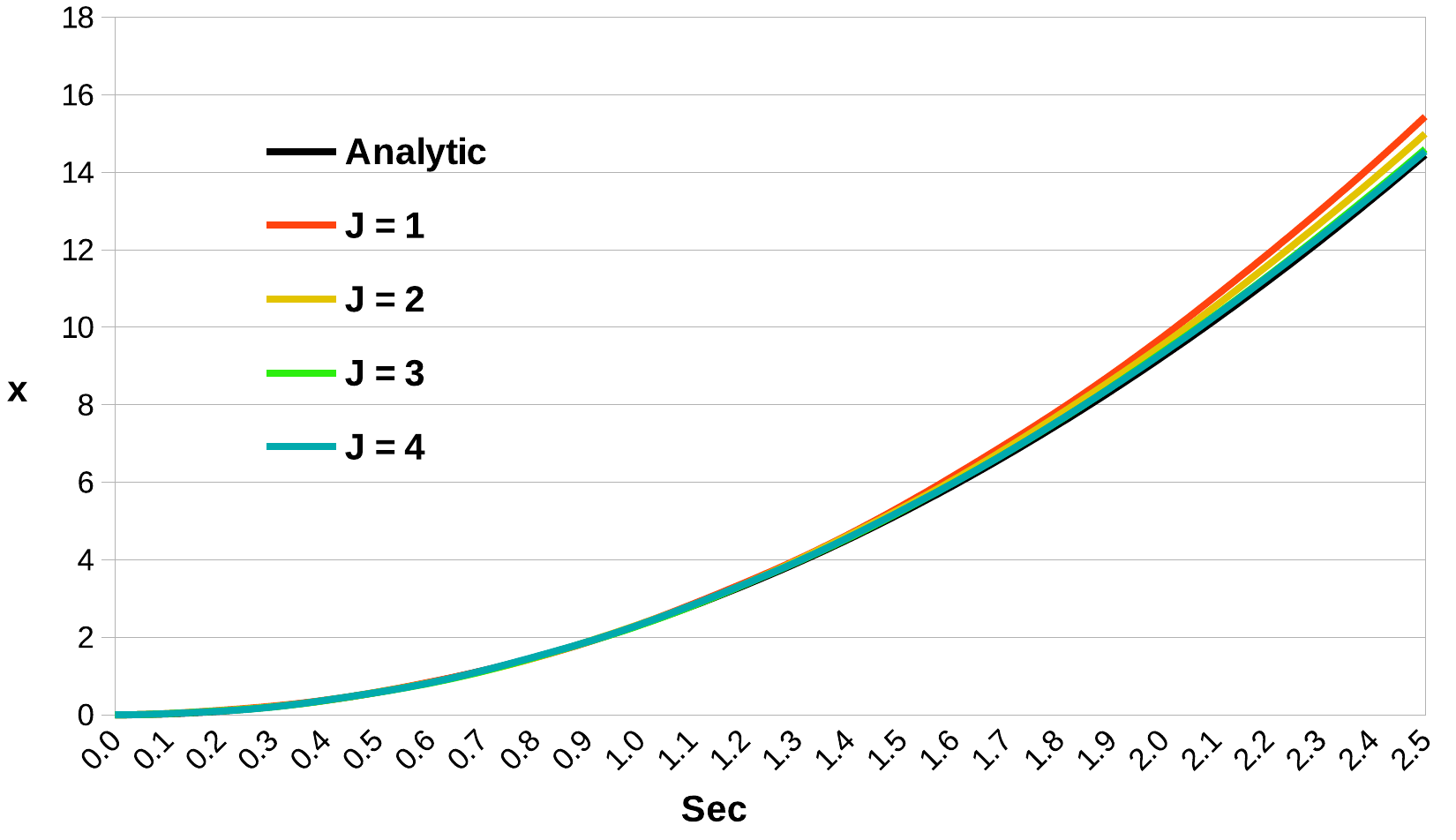}\\
II) \includegraphics[scale=0.55]{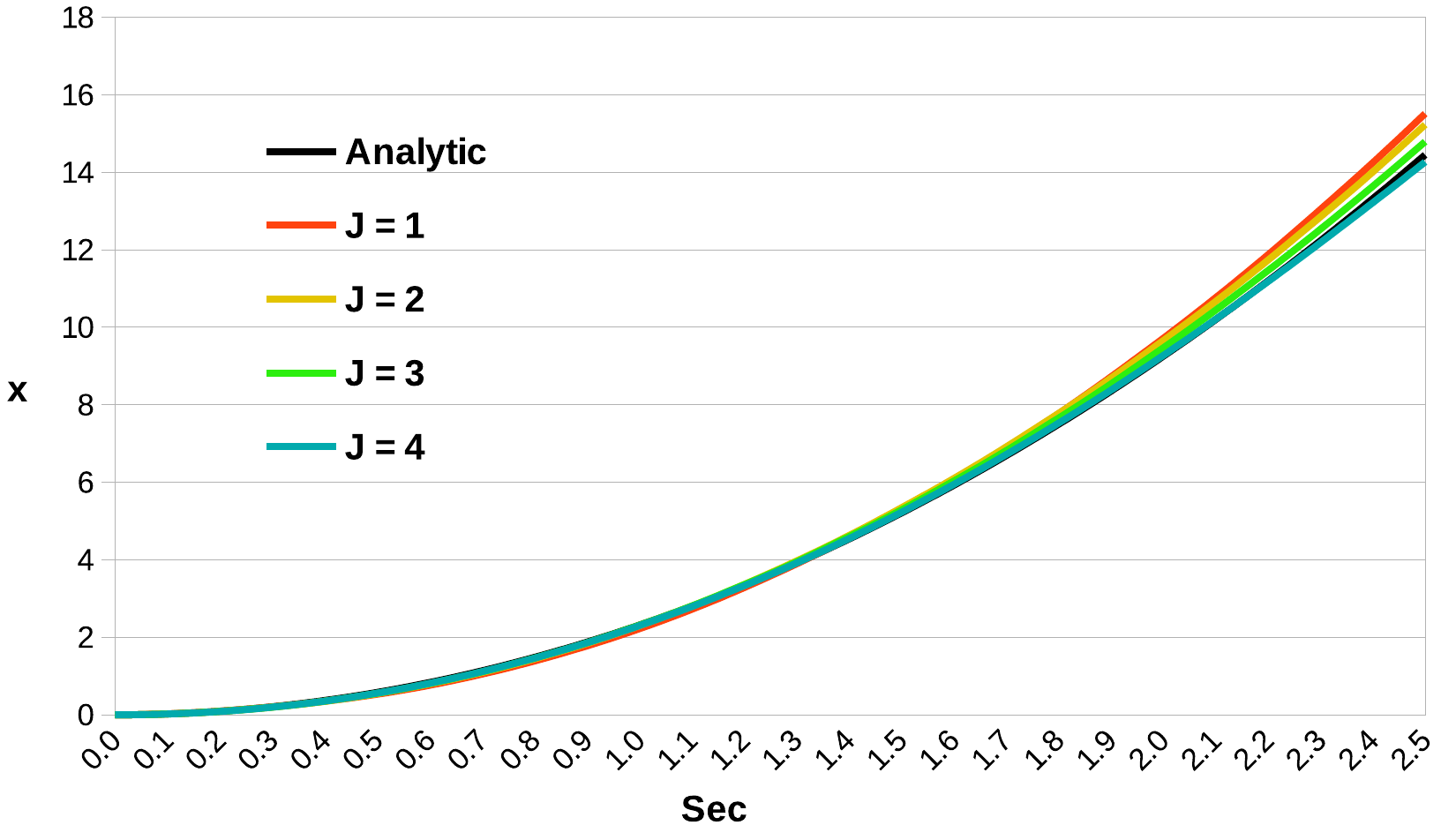}
\caption{\giachi{Position of the center of mass of the rolling disk. I): $E = E_1 = 4.2 \,\cdot\, 10^6 \, Pa$. II): $E = E_2 = 4.2 \,\cdot\, 10^8 \, Pa$.}}\label{CE}
\end{figure}

 We begin by considering a 2-dimensional example of a disk rolling on an inclined plane.
 The case of a rolling disk (or ball in 3D) is a common test for the MPM, and it has been used in several other works, we report for instance \cite{bardenhagen2000material, chen2015improved, lian2011coupling}.
 The plane has an inclination of $\theta = \pi / 4$, the disk has radius $R=1.6 \,m$, and Young's modulus $E$.
 Two values of $E$ are chosen for the simulations, $E_1 = 4.2 \,\cdot\, 10^6 \, Pa$ and $E_2 = 4.2 \,\cdot\, 10^8 \, Pa$.  
 \giachi{The coarse mesh is composed of $4$ horizontal layers, each with $20$ bi-quadratic quadrilateral elements, 
 so each of them has unitary length in the $x$-direction. 
 Because the mesh is progressively refined moving towards the plane, each layer has a different cell size in the $y$-direction. These lengths are $1.72 \, m$, $1.34 \, m$, $1.09 \, m$, and $0.85 \, m$.
 Finer grids are obtained with midpoint refinement. Hence, after every refinement, the number of elements increases by a factor of four. Let $J$ denote the number of refinements carried out for a given simulation. For instance, if $J=3$, the coarse background grid is refined $3$ times. If $J=0$, it means that the coarse grid is not refined. The values of $J=1,2,3,4$ are chosen for the tests. 
 The initial distance $d_0(J)$ of the center of the disk from the plane depends on the refinement of the grid, 
 and it is designed to keep a small gap between the disk and the plane. For given $J$ and $E$, the gap is a fraction of the height of the element of the background grid where the contact occurs.  
 We choose $d_0(J)=R + 0.2 / 2^{J-1}$ for $E_1$, and $d_0(J) = R + 0.3 / 2^{J-1}$ for $E_2$.}
 The reference frame is translated and centered at the center of the disk as in Figure \ref{rolling}. 
 The analytic expression of the $x$-coordinate of the center of mass assuming an undeformable disk is given by
 \begin{align}\label{analytic}
  x(t) = x_0 + \dfrac{1}{3} \, g \, t^2 \, sin(\theta),
 \end{align}
where $x_0$ is assumed to be zero.
 Concerning the other simulation parameters, the density of the disk is $1000 \, kg/m^3$, the Poisson's coefficient is $\nu = 0.4$, and the step size is $\Delta t = 0.01 \, s$. For Newmark's \giachi{integrator}, $\beta= 0.3$ and $\gamma = 0.5$ are selected.
 Rolling without slip is considered \giachi{and $C_c = 0.5$.} 
\giachi{
% Such features have been implemented to provide an accurate capture of the rolling of the disk on the plane.
% As a matter of fact, a precise modeling of the interaction between the disk and the plane is key to obtain accurate results
The input parameters for the generation of the disk are $(x_c,y_c)=(0,0)$, $R_0=1.4\,m$, $R=1.6\,m$, $N_{\theta_0} = 300$, and $N_b = 22$, for a total of $48739$ particles.
The pseudocode for the initialization of the MPM disk can be found in Appendix A.
In Figure \ref{rolling_mat}, the particle distribution (including the particle boundary layers) and some details about the graded mesh for the background grid are visible.
In Figure \ref{rolling_mat} the values of the flags $M_{\mathcal{E}}$ are also reported for the elements of the background grid.
Recall that since bi-quadratic elements are considered, the maximum possible value of $M_{\mathcal{E}}$ is 9, which is attained for  interior elements.}
 \begin{figure}[t]
\centering
\includegraphics[scale=0.43]{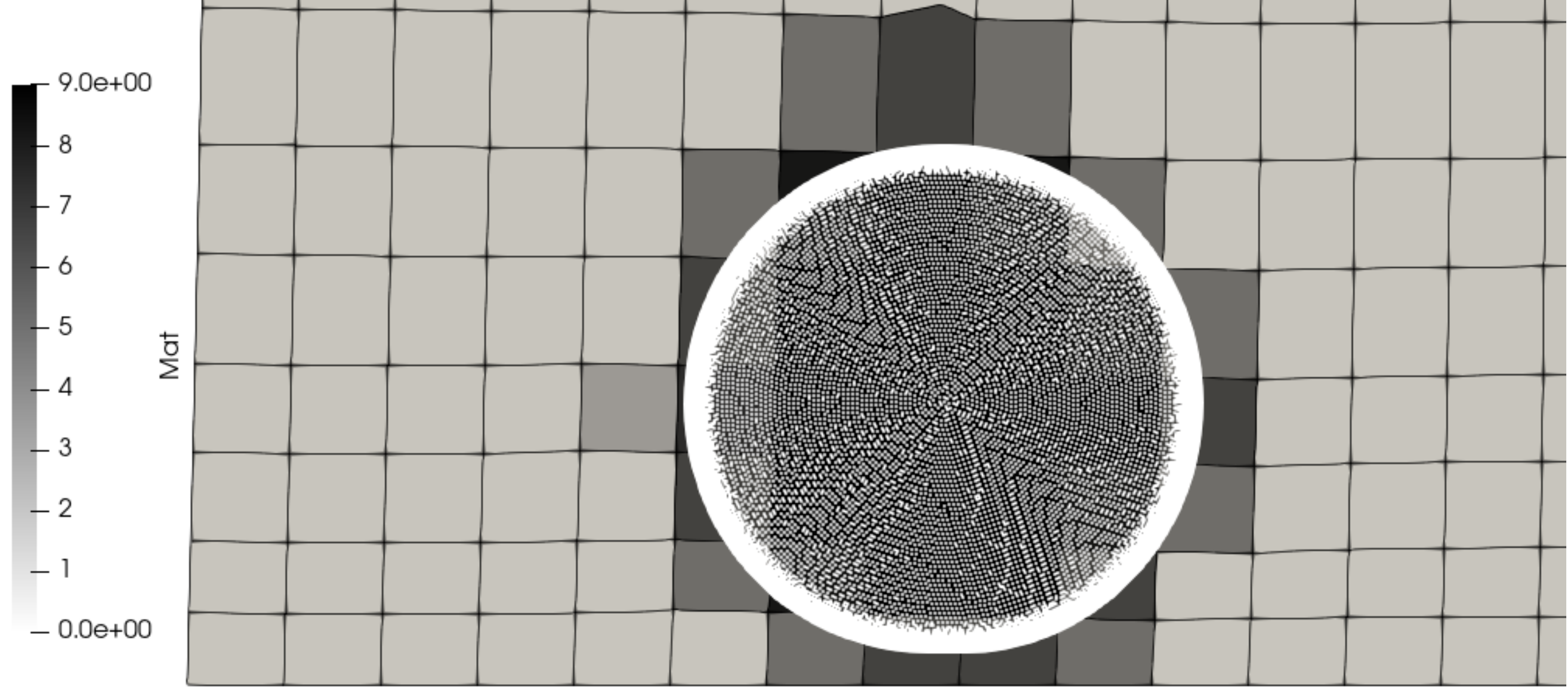}
\caption{Rolling disk simulation for $E = E_1 = 4.2 \,\cdot\, 10^6 \, Pa$. The different shades of gray for the elements of the background grid identify different values of $M_{\mathcal{E}}$, which is referred to as Mat in the legend. One refinement ($J=1$) has been considered for the background grid.}\label{rolling_mat}
\end{figure}
The lightest gray refers to the elements of the inactive background grid. The following values were chosen for the soft stiffness scaling factor $C_{\mathcal{E}}$ in Eq. \eqref{softResidualLocal}
\giachi{
\begin{equation}\label{CE1}
 \begin{aligned} \mbox{For $E_1$:} \,\, C_{\mathcal{E}} = 
 \begin{cases}
  10^{-3} & \mbox{if} \quad M_{\mathcal{E}} < 5, \\
  10^{-7} & \mbox{if} \quad 5 \leq M_{\mathcal{E}} < 9,\\
  0 & \mbox{if} \quad  M_{\mathcal{E}} = 9,
 \end{cases}
 \end{aligned}
 \end{equation}
 \begin{equation}\label{CE2}
  \begin{aligned} \mbox{For $E_2$:} \,\, C_{\mathcal{E}} = 
 \begin{cases}
  10^{-5} & \mbox{if} \quad M_{\mathcal{E}} < 5, \\
  10^{-9} & \mbox{if} \quad 5 \leq M_{\mathcal{E}} < 9,\\
  0 & \mbox{if} \quad  M_{\mathcal{E}} = 9.
 \end{cases}
 \end{aligned}
\end{equation}}
The horizontal position of the center of mass obtained numerically solving our implicit MPM scheme is compared to the analytic expression in Eq. \eqref{analytic} for different values of $J$, and $E$.
Results for $E_1$ are visible in Figure \ref{CE} I), for different mesh sizes.
The curves obtained with the proposed implicit MPM method are in 
agreement with the analytic equation of the center of mass in Eq. \eqref{analytic}. 
\giachi{It is also shown that a progressive refinement of the mesh provides increasingly accurate results, since the curve associated with $J = 4$ is the one that best overlaps with the analytic expression.
Results for $E_2$ are reported in Figure \ref{CE} II). Once again, the curves are in 
agreement with the analytic expression and larger values of $J$ give better accuracy. }

\giachi{Average relative errors are also shown, to provide a quantitative measure of the error. If $x(t)$ represents the position of the center of mass defined in Eq. \eqref{analytic}, let $\overline{x}(t)$ be the position computed with the proposed method.
Then the averaged relative error is defined as
\begin{align}
 \dfrac{1}{N_t}\sum\limits_{i=1}^{N_t} \dfrac{|x(t_i) - \overline{x}(t_i)|}{x(t_i)},
\end{align}
where $N_t=250$ is the total number of time steps considered.
Results for the above error are reported in Table \ref{errEx1} corresponding to the curves in Figure \ref{CE}.}

\begin {table}[!t]
\setlength\tabcolsep{3.25pt} % default value: 6pt
\begin{center}                                                               
	\begin{tabular}{|c|c|c|c|c|c|} \hline	
% 	 \multicolumn{7}{|c|}{ $N_{\gamma} = 2 $  } \\ \cline {1-7}   
% 	  &  \multicolumn{2}{|c|}{ N=1 } &  \multicolumn{2}{|c|}{ N=2 } &  \multicolumn{2}{|c|}{ N=3 }\\ \cline {2-7}  
  $E$   & $C_{\mathcal{E}}$   &   $J = 1$ &   $J = 2$  &  $J = 3$   &   $J = 4$    \\ \hline 
  $4.2 \cdot 10^6$  & Eq. \eqref{CE1}     & 5.545e-02 &  3.921e-02&  2.571e-02 &   2.253e-02 \\ \hline
  $4.2 \cdot 10^8$  & Eq. \eqref{CE2}     & 4.520e-02 &  3.502e-02&  2.423e-02 &   1.720e-02  \\ \hline
  	\end{tabular}
\end{center}
\caption{\giachi{averaged relative error associated with the results in Figure \ref{CE}.}}
\label{errEx1}
\end{table}

\begin{figure}[htb]
\centering
\includegraphics[scale=0.21]{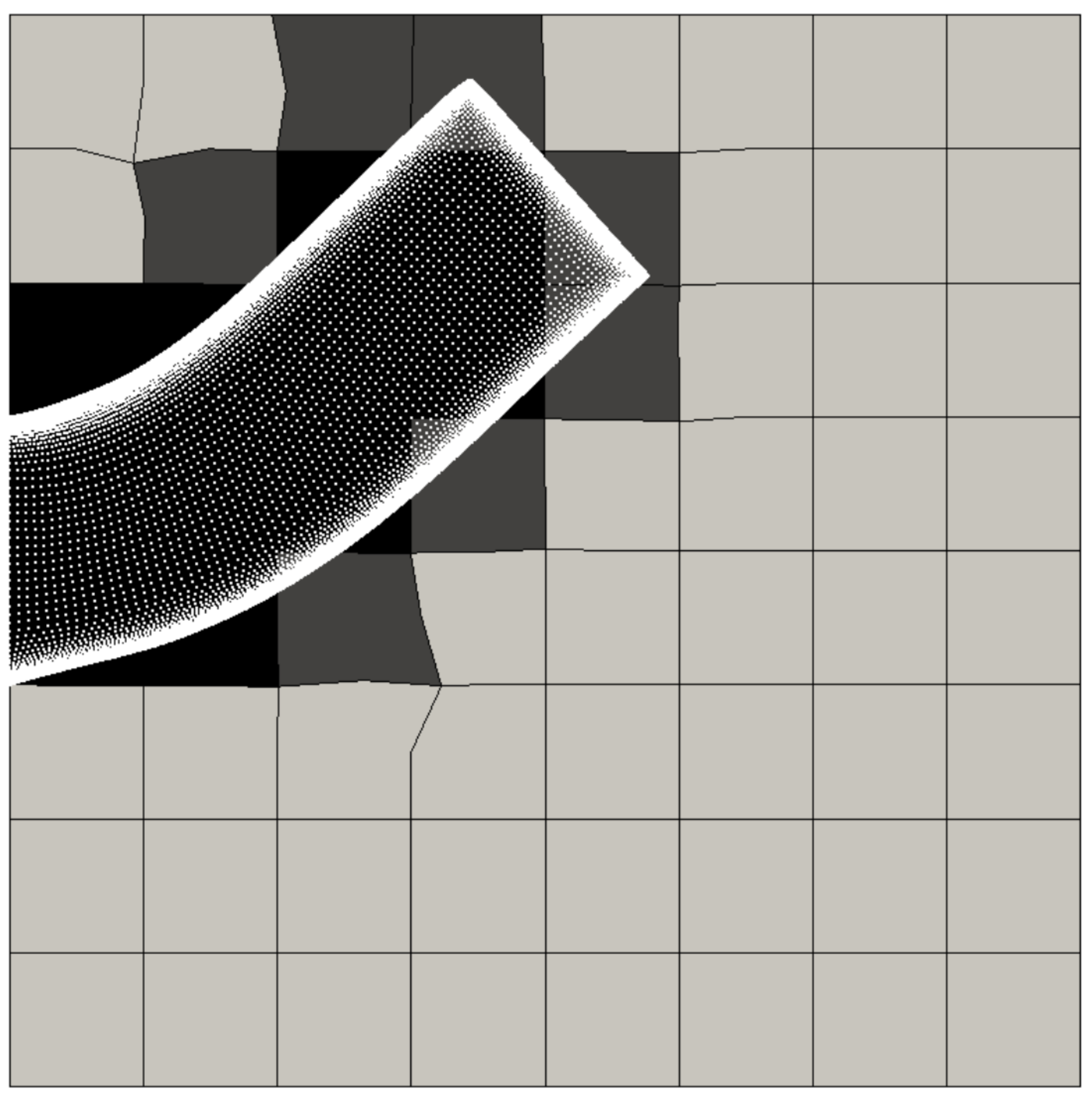}
\includegraphics[scale=0.21]{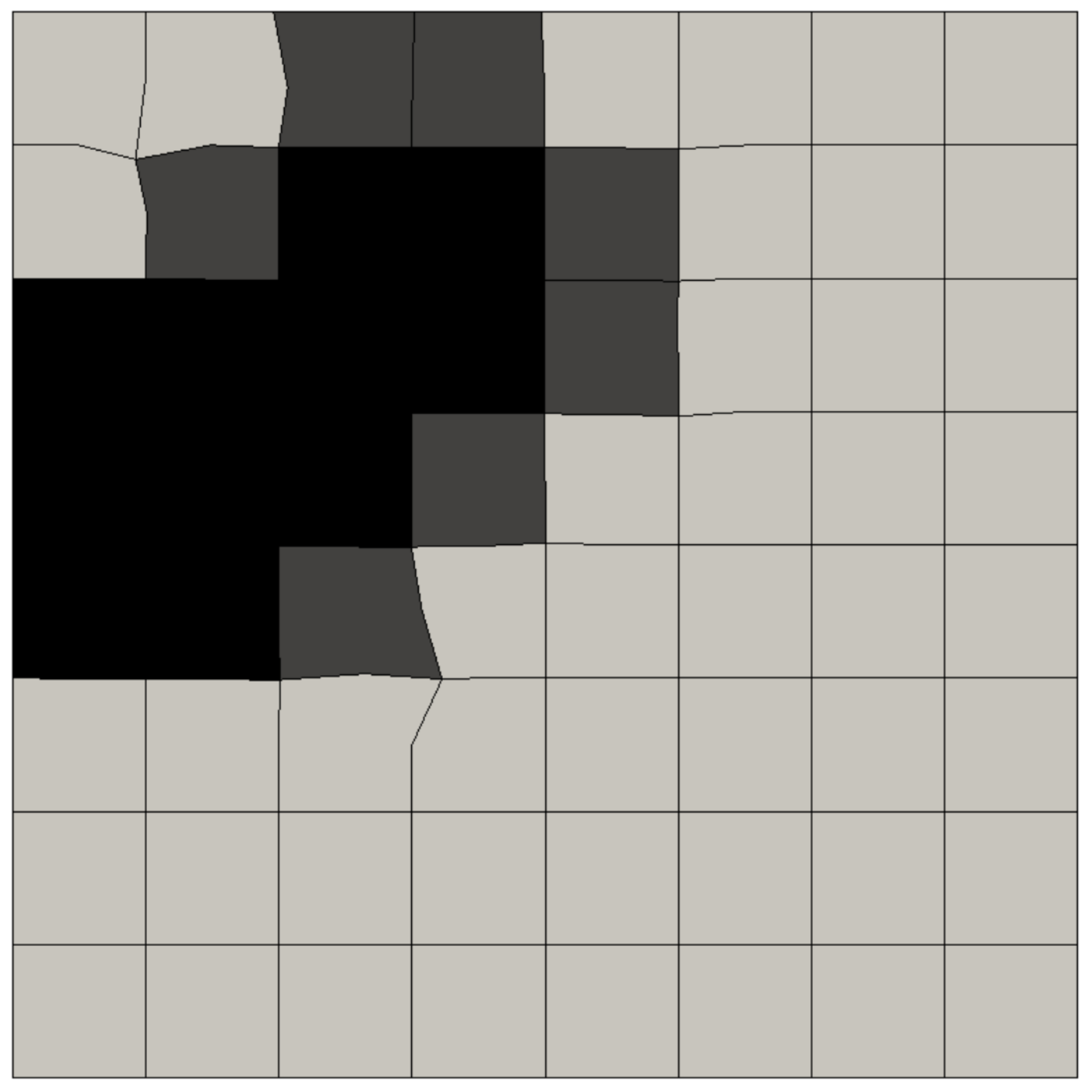}
\caption{Cantilever beam simulation with $J=2$. The MPM particles are omitted on the right figure to highlight the deformations and the values of $M_{\mathcal{E}}$ for the background grid elements.}\label{beam}
\end{figure}

\begin{figure}[t!]
\centering
\includegraphics[scale=0.55]{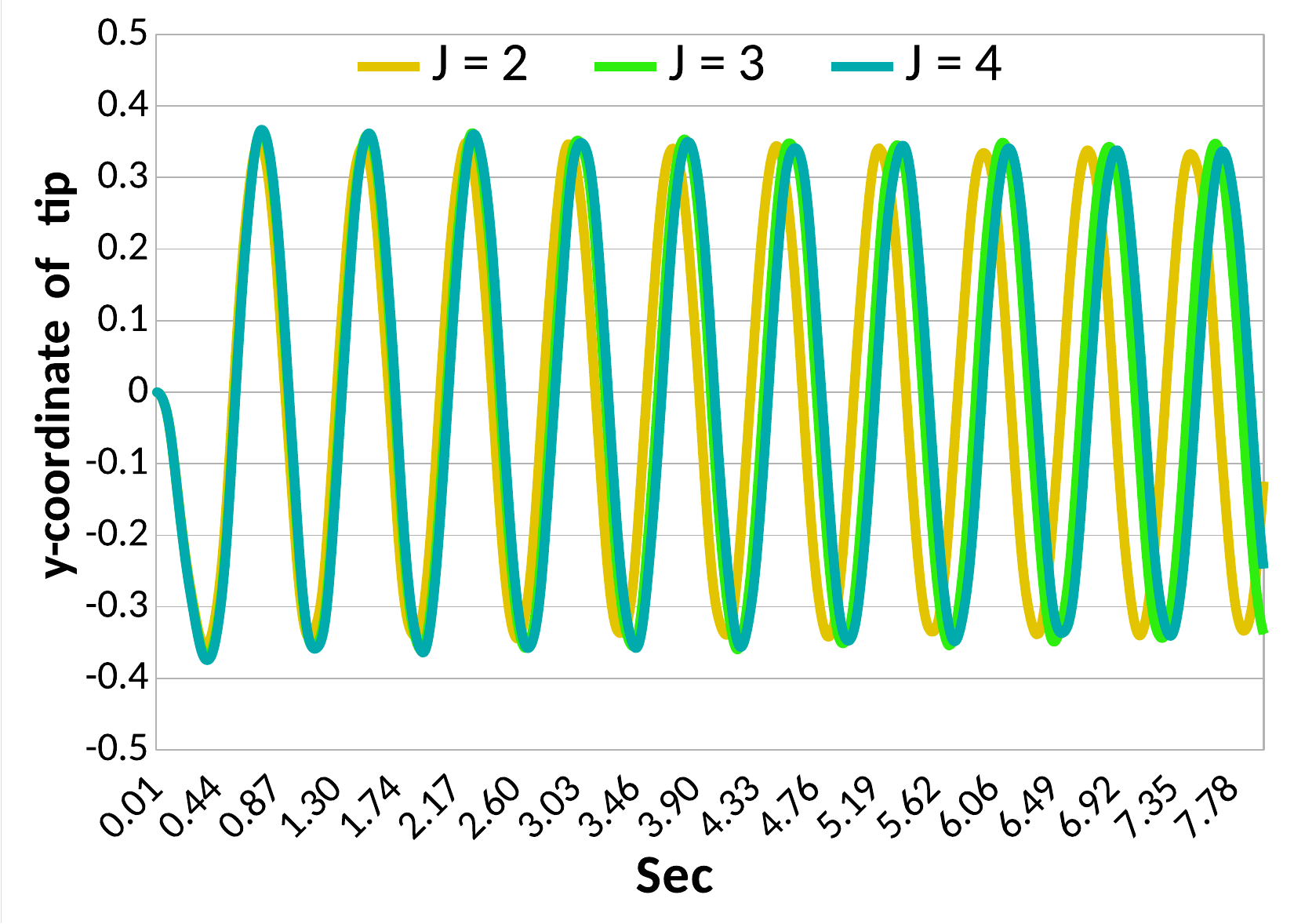}
\caption{Oscillations of the tip of the cantilever beam, as the resolution of the background grid is varied.}\label{beamResults}
\end{figure}

The section continues with simulations of a cantilever beam with length $L = 0.625 \, m$ and height $H = 0.25 \, m$, visible in Figure \ref{beam}.
\giachi{The parameters to build the MPM beam are $(x_0,y_0) =(0,0)$, $L = 0.625 \, m$, $H = 0.25 \, m$, $H_0 = 0.15 \, m$, $N_H=20$, and $N_b=21$, for a total of $23041$ particles.
The pseudocode for the MPM beam initialization is given in Appendix A. 
The domain is a unit box, whose center of the left boundary is placed at the origin,
and the coarse mesh is composed of 4 bi-quadratic quadrilateral elements.}
The background grid obtained when $J=2$ is visible in Figure \ref{beam}.
The beam has density $\rho = 10,000 \, kg/m^3$, Poisson's coefficient $\nu = 0.4$, and Young's modulus $E = 1.74 \, \cdot \, 10^6 \, Pa$. The step size is $\Delta t = 0.008 \, s$ and, as before, $\beta = 0.3$ and $\gamma = 0.5$ are chosen for Newmark's \giachi{integrator}.
\giachi{The values of $C_{\mathcal{E}}$ are
 \begin{equation}\label{CE3a}
  \begin{aligned} \mbox{For $E_2$:} \,\, C_{\mathcal{E}} = 
 \begin{cases}
  10^{-6} & \mbox{if} \quad M_{\mathcal{E}} < 5, \\
  10^{-10} & \mbox{if} \quad 5 \leq M_{\mathcal{E}} < 9,\\
  0 & \mbox{if} \quad  M_{\mathcal{E}} = 9.
 \end{cases}
 \end{aligned}
\end{equation}
}
The simulation is carried out in the following way: the beam is initially subject to its self weight, so only the gravitational 
force is applied to it. Then, when it reaches its larges deformation, is assumed that the gravitational force is removed and the oscillations of the tip of the beam are monitored.
If no significant damping in the oscillations can be observed as time increases, it means that the method is accurate.
Graphs for such oscillations are reported in Figure \ref{beamResults}, for values of $J=2$, $J=3$ and $J=4$.
From Figure \ref{beamResults} it can be seen that, as time increases, the oscillations became slightly off-phase as the mesh becomes finer.
Although, the damping of the oscillations is very small, for all mesh sizes considered, despite the long duration of the simulation which is 8 seconds. 
%The numerical results reported show how the implicit MPM method proposed in this work achieves great accuracy.
 
 \subsection{Tests for the MPM-FEM coupling}
 
\begin{figure}[!t]
\centering
I) \includegraphics[scale=0.55]{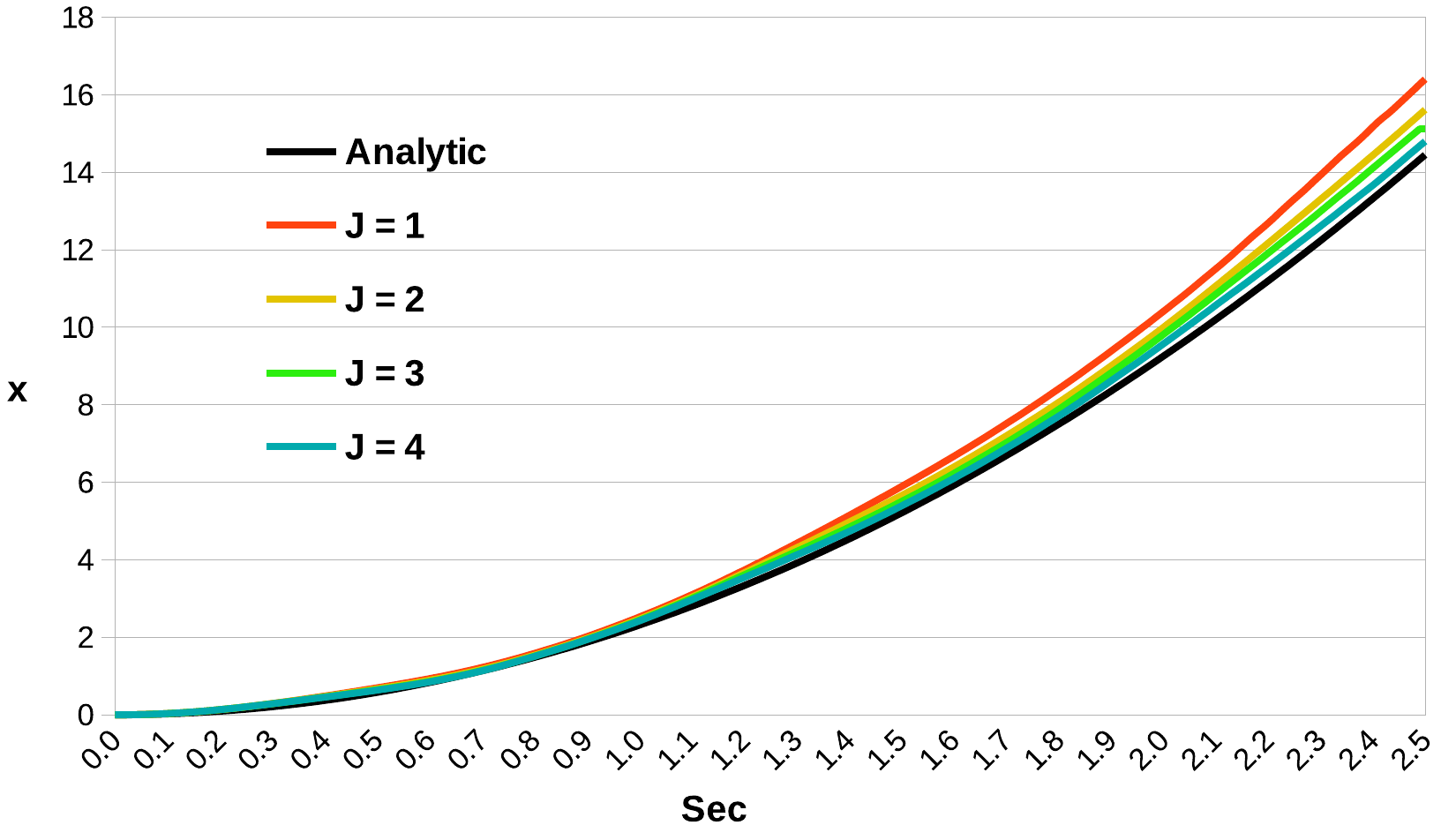} \\
II) \includegraphics[scale=0.55]{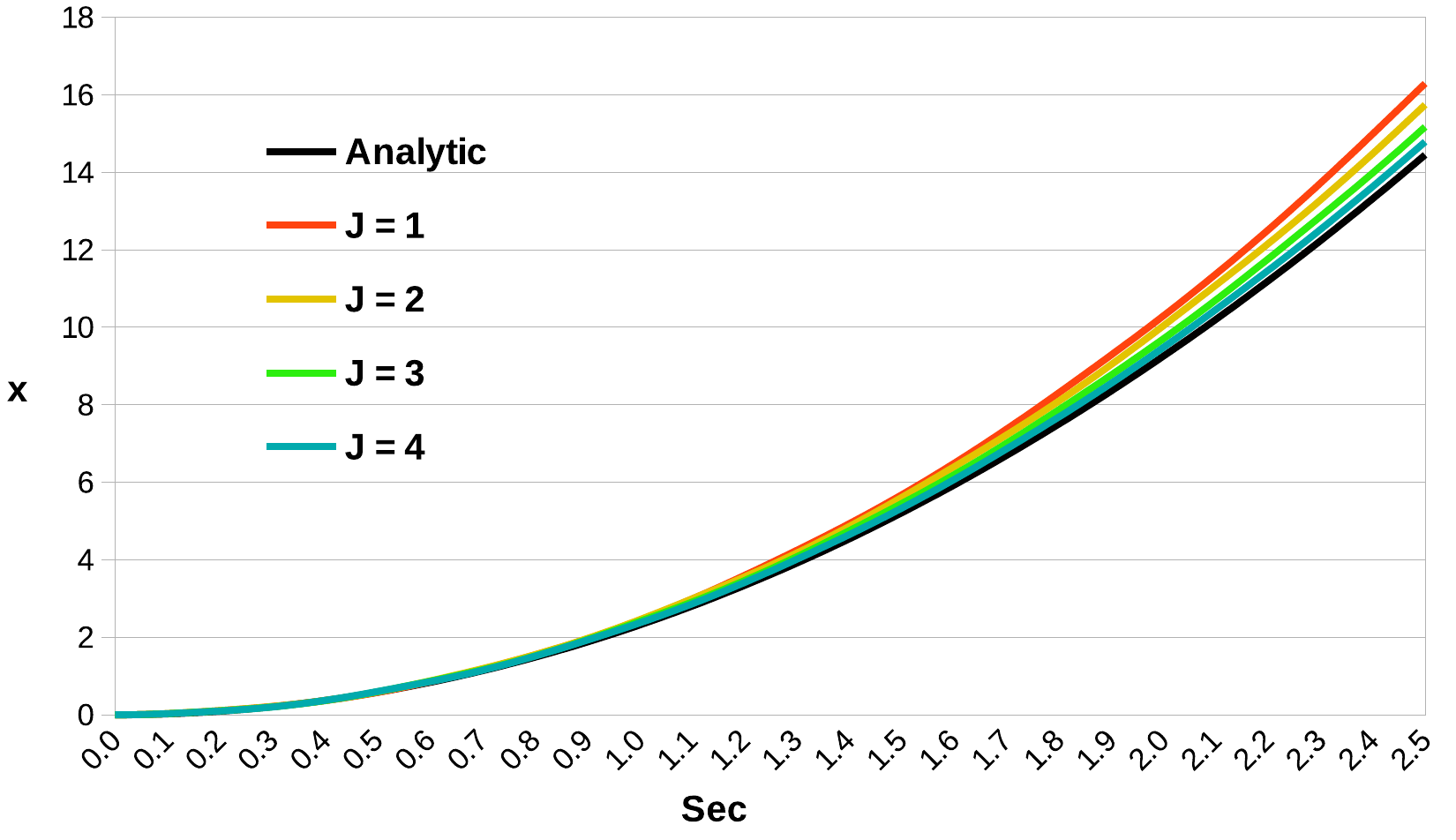} \\
III) \includegraphics[scale=0.55]{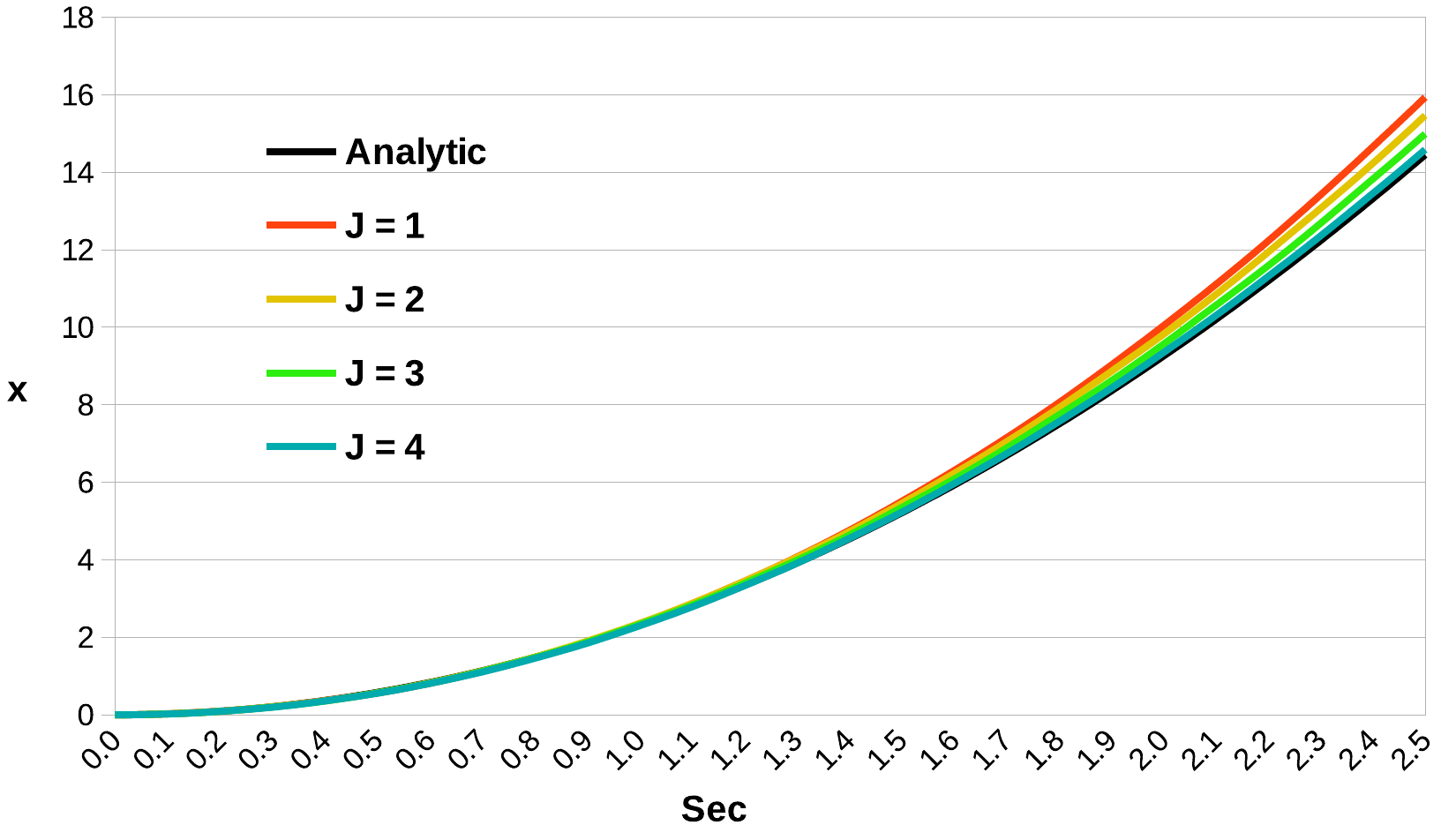}
\caption{Position of the center of mass of the rolling disk for the coupled MPM-FEM case. I): $E_{mpm} = 4.2 \cdot 10^6\, Pa$ and $E_{fem} = 8.4 \cdot 10^6 \, Pa$. II): $E_{mpm} = 4.2 \cdot 10^7\, Pa$ and $E_{fem} = 8.4 \cdot 10^7 \, Pa$. III): $E_{mpm} = 4.2 \cdot 10^8\, Pa$ and $E_{fem} = 8.4 \cdot 10^8 \, Pa$. }\label{ex3_tests}
\end{figure}

\begin{figure}[t!]
 \centering
\includegraphics[scale=0.4]{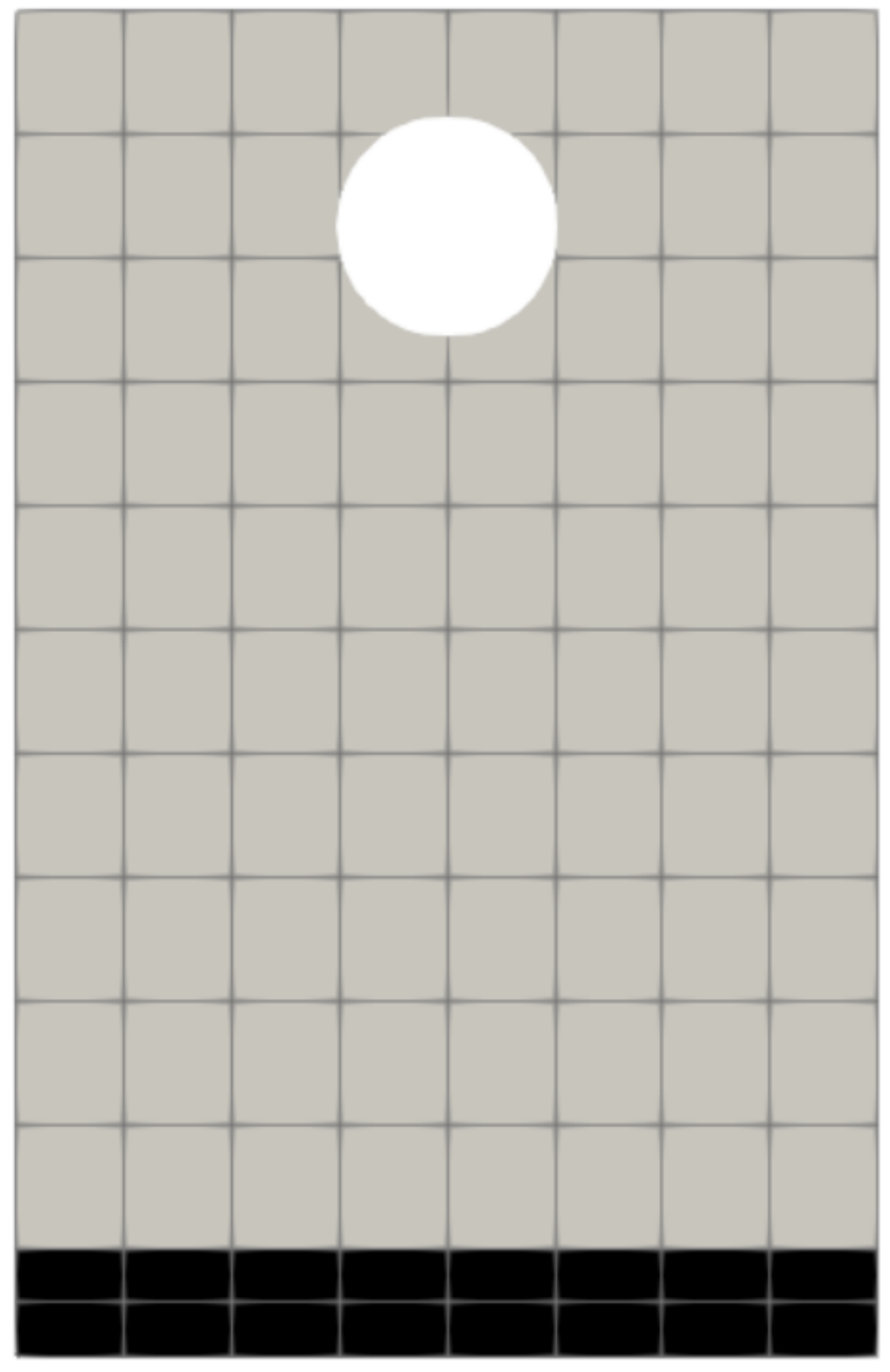}
\caption{Disk falling on a horizontal plate. The disk (white) is modeled with MPM, and the plate (black) with FEM. The coarse background grid is in gray color.}\label{basket}
\end{figure}

\begin{figure}[t!]
 \centering
\includegraphics[scale=0.2]{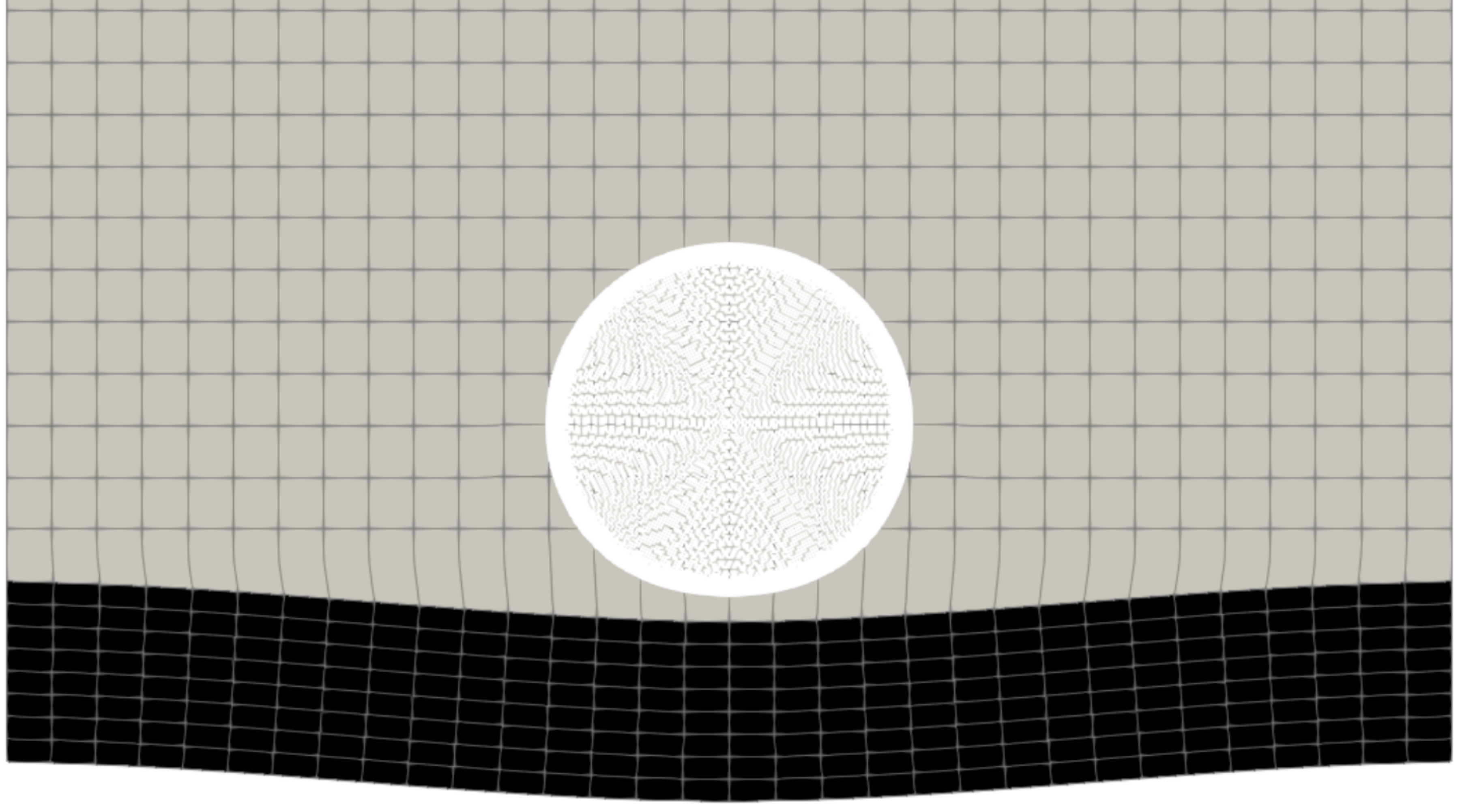}
\caption{Deformation of the disk and plate as the disk impacts on the plate.}\label{basket2}
\end{figure}

 In this section, numerical results for the MPM-FEM coupling are presented. We begin by considering a disk rolling on an inclined plate, as in \cite{lian2011coupling, chen2015improved}.
 The disk is modeled with MPM while the plate is discretized with FEM. The dimensions of the plate are $20 \times 1.6 \, m$ while the radius of the disk is $R = 1.6 \, m$. Rolling without slip is considered and the inclination of the plate is $\theta = \pi / 4$.  
   \giachi{The MPM background grid is the same as the one used for the rolling ball in the uncoupled case, whereas for the FEM plate two horizontal layers of $20$ elements each are considered, with $x$ length $1 \, m$ and $y$ length $0.85 \, m$.} 
  Recall that the use of a background grid for the MPM attached to the FEM body is one of the major contributions of the present work. \giachi{The coupling coefficient $C_c$ is equal to $0.5$.} 
  The coarse grid is refined by means of midpoint refinement, with $J$ indicating the number of refinements performed.
   The density and the Poisson's coefficient of both the disk and the plate are set to $1000 \, kg/m^3$ and $\nu = 0.4$, respectively.
 For what concerns the time integration,  $\Delta t = 0.01 \, s$, and the parameters for Newmark's \giachi{integrator} are $\beta= 0.3$ and $\gamma = 0.5$.
  \giachi{The Young's moduli of the MPM and the FEM bodies are respectively $E_{mpm} = 4.2 \cdot 10^{\alpha}\, Pa$ and $E_{fem} = 8.4 \cdot 10^{\alpha} \, Pa$, with  $\alpha =6,7,8$.
 The initial vertical distance $d_0(J)$ of the center of the disk from the plane is 
 $d_0(J)= R + 0.15 / 2^{J-1}$ for $\alpha=6$, 
 $d_0(J)= R + 0.2  / 2^{J-1}$ for $\alpha=7$, and 
 $d_0(J)=R + 0.25 / 2^{J-1}$ for $\alpha=8$.
Again this is done so that there exists an initial small gap between the disk and the plane, 
and it is designed so that the gap dimension is a fraction of the height of the background grid element where the contact occurs.}
 The reference frame is then translated and centered at the center of mass of the disk, hence the analytic expression of the $x$-coordinate of the center of mass is given in Eq. \eqref{analytic}. 
\giachi{The parameters for the initialization of the MPM disk are now $(x_c,\,y_c) = (0,\,0)$, 
$R = 0.16 \, m$, $R_0=0.14 m$, $N_{\theta_0}=600$ and $N_b = 24$, for a total of $57084$ particles.
The coefficient $C_{\mathcal{E}}$ for the present tests has been chosen as follows:
\begin{equation}\label{alpha6}
 \begin{aligned} \mbox{For $\alpha=6$:} \,\, C_{\mathcal{E}} = 
 \begin{cases}
  10^{-2} & \mbox{if} \quad M_{\mathcal{E}} < 5, \\
  10^{-6} & \mbox{if} \quad 5 \leq M_{\mathcal{E}} < 9,\\
  0 & \mbox{if} \quad  M_{\mathcal{E}} = 9,
 \end{cases}
 \end{aligned}
 \end{equation}
 \begin{equation}\label{alpha7}
  \begin{aligned} \mbox{For $\alpha=7$:} \,\, C_{\mathcal{E}} = 
 \begin{cases}
  10^{-3} & \mbox{if} \quad M_{\mathcal{E}} < 5, \\
  10^{-7} & \mbox{if} \quad 5 \leq M_{\mathcal{E}} < 9,\\
  0 & \mbox{if} \quad  M_{\mathcal{E}} = 9.
 \end{cases}
 \end{aligned}
\end{equation}
 \begin{equation}\label{alpha8}
  \begin{aligned} \mbox{For $\alpha=8$:} \,\, C_{\mathcal{E}} = 
 \begin{cases}
  10^{-4} & \mbox{if} \quad M_{\mathcal{E}} < 5, \\
  10^{-8} & \mbox{if} \quad 5 \leq M_{\mathcal{E}} < 9,\\
  0 & \mbox{if} \quad  M_{\mathcal{E}} = 9.
 \end{cases}
 \end{aligned}
\end{equation}
}
%%%%%%%%%%%%%%%%%%%%%%%%%%%%%%%%%%%%%%%%%%%%%%%%%%%%%%%%%%%%%%%%%%%%%%%%%%%%%%%%%%%%%%%%%%%%%%%%%%%%%%%%%%%%%%%%%%%%%%%%%%%%%%%%%
\begin {table}[!t]
\setlength\tabcolsep{3.25pt} % default value: 6pt
\begin{center}                                                               
	\begin{tabular}{|c|c|c|c|c|c|c|} \hline	
% 	 \multicolumn{7}{|c|}{ $N_{\gamma} = 2 $  } \\ \cline {1-7}   
% 	  &  \multicolumn{2}{|c|}{ N=1 } &  \multicolumn{2}{|c|}{ N=2 } &  \multicolumn{2}{|c|}{ N=3 }\\ \cline {2-7}  
  $E_{mpm}$   & $E_{fem}$ &$C_{\mathcal{E}}$   &   $J = 1$ &   $J = 2$  &  $J = 3$   &   $J = 4$    \\ \hline 
  $4.2 \cdot 10^5$  &  $8.4 \cdot 10^5$ & Eq. \eqref{alpha6}  & 1.604e-01 &  1.263e-01&  1.051e-01 &   8.783e-02 \\ \hline
  $4.2 \cdot 10^6$  &  $8.4 \cdot 10^6$ & Eq. \eqref{alpha7}  & 9.029e-02 &  7.918e-02 & 5.416e-02 &   3.654e-02  \\ \hline
  $4.2 \cdot 10^7$  &  $8.4 \cdot 10^7$ & Eq. \eqref{alpha8}  & 5.368e-02 &  4.289e-02 & 2.919e-02 &   1.959e-02  \\ \hline
  	\end{tabular}
\end{center}
\caption{\giachi{averaged relative error associated with the results in Figure \ref{ex3_tests}.}}
\label{errEx3}
\end{table}
%%%%%%%%%%%%%%%%%%%%%%%%%%%%%%%%%%%%%%%%%%%%%%%%%%%%%%%%%%%%%%%%%%%%%%%%%%%%%%%%%%%%%%%%%%%%%%%%%%%%%%%%%%%%%%%%%%%%%%%%%%%%%%%%%
\giachi{In Figure \ref{ex3_tests}, results are displayed for $\alpha =6,7,8$}. 
We observe that also in the coupled case there is agreement between the numerical results and the analytic position of the center of mass.
\giachi{Once again, increasing the value of $J$ provides better accuracy with respect to the analytic curve.
Moreover, to stiffer materials it corresponds a smaller error. This is also expected since the analytic solution
is obtained under the assumption of an undeformable body.
The averaged relative errors reported in Table \ref{errEx3} confirm the accuracy and validity of the proposed coupling method. }
% \eugenio{The plate deformation can be seen clearly, for instance, in Figure \ref{def55}.}
%
 %
%  \begin{figure}[htb]
%  \centering
% \includegraphics[scale=0.35]{Figures/coupled_disk_3ref5e5}
% \caption{Deformation of the rolling disk and plate.}\label{def55}
% \end{figure}
%
This section is concluded with simulations of a disk falling from above and bouncing on a horizontal plate. As before, the disk is modeled with MPM and the plate is discretized with FEM. 
\giachi{The coarse MPM background grid has dimensions $1.28 \times 1.84 \, m$ and it is made of 80 elements of $x$ length $0.16 \, m$ and $y$ length $0.184 \, m$. The FEM plate has dimensions $1.28 \times 0.64 \,m$ and is discretized with 2 horizontal layers of 8 cells each, of $x$ length $0.16 \,m$ and $y$ length $0.08 \,m$.}
The disk has radius $R = 0.16 \, m$ and is placed initially 
at the center-top of the background grid such that its distance from the upper surface of the plate is $1.68 \,m$.
Schematics of the coarse background grid and discretizations are visible in Figure \ref{basket}, while a picture of the impact between the disk and the plate can be found in Figure \ref{basket2}
\giachi{
The parameters for the initialization of the MPM disk are now $(x_c,\,y_c) = (0,\,0)$, 
$R = 0.16 \, m$, $R_0=0.14 m$, $N_{\theta_0}=300$ and $N_b = 22$, for a total of $48739$ particles. 
For what concerns the physical parameters,  both the disk and the plate have density $ \rho_{mpm} =  \rho_{fem} = 1000 \, kg/m^3$ and Poisson's coefficient  $\nu = 0.4$.} The Young's moduli for the MPM and FEM body are, respectively, $E_{mpm} = 5.91 \cdot 10^6 \, Pa$ and $E_{fem} = 4.2 \cdot 10^7 \, Pa$.
The aim of these tests is to further demonstrate the accuracy of the proposed methods and to highlight how it improves as the values of the coefficient
$C_{\mathcal{E}}$ is decreased, considering fixed values of $E$.
This behavior is explained by the fact that for fixed Young's modulus, smaller values of $C_{\mathcal{E}}$ decrease the soft stiffness contribution and therefore decrease the error added to the method.
\giachi{The values of $C_{\mathcal{E}}$ for this example are as follows
\begin{equation}\label{CE3}
 \begin{aligned}  C_{{\mathcal{E}},1} = 
 \begin{cases}
  10^{-2} & \mbox{if} \quad M_{\mathcal{E}} < 5, \\
  10^{-6} & \mbox{if} \quad 5 \leq M_{\mathcal{E}} < 9,\\
  0 & \mbox{if} \quad  M_{\mathcal{E}} = 9,
 \end{cases}
 \end{aligned}
 \end{equation}
 \begin{equation}\label{CE4}
  \begin{aligned}  C_{{\mathcal{E}},2} = 
 \begin{cases}
  10^{-5} & \mbox{if} \quad M_{\mathcal{E}} < 5, \\
  10^{-9} & \mbox{if} \quad 5 \leq M_{\mathcal{E}} < 9,\\
  0 & \mbox{if} \quad  M_{\mathcal{E}} = 9.
 \end{cases}
 \end{aligned}
\end{equation}}
\giachi{The value of the coupling coefficient $C_c$ depends on the refinement level and on $C_{\mathcal{E}}$ in this case.
Its values are reported in Table \ref{Cc}.
To speed up the simulation, the time stepping is adaptive, meaning that $\Delta t$ is large when the disk is far from the plate and
it is suddenly decreased after the disk reaches a given distance from the plate. The value of $\Delta t$ when the disk is near the plate is reported in Table \ref{Cc}.}
The parameters for Newmark's \giachi{integrator} are once again $\beta= 0.3$ and $\gamma = 0.5$.
%%%%%%%%%%%%%%%%%%%%%%%%%%%%%%%%%%%%%%%%%%%%%%%%%%%%%%%%%%%%%%%%%%%%%%%%%%%%%%%%%%%%%%%%%%%%%%%%%%%%%%%%%%%%%%%%%%%%%%%%%%%%%%%%%
\begin {table}[!t]
\setlength\tabcolsep{3.25pt} % default value: 6pt
\begin{center}                                                               
	\begin{tabular}{|c|c|c|c|} \hline	
% 	 \multicolumn{7}{|c|}{ $N_{\gamma} = 2 $  } \\ \cline {1-7}   
% 	  &  \multicolumn{2}{|c|}{ N=1 } &  \multicolumn{2}{|c|}{ N=2 } &  \multicolumn{2}{|c|}{ N=3 }\\ \cline {2-7}  
\multicolumn{4}{|c|}{ $ C_c $  } \\ \cline {1-4} 
                     &  $J = 2$  &  $J = 3$   &   $J = 4$    \\ \hline 
 $C_{\mathcal{E},1}$   & 0.06&  0.1 &  0.2 \\ \hline
 $C_{\mathcal{E},2}$   & 0 & 0.1 & 0.2  \\ \hline
 \multicolumn{4}{|c|}{ $ \Delta t_{plate} \, (sec)$   } \\ \cline {1-4} 
                     &  $J = 2$  &  $J = 3$   &   $J = 4$    \\ \hline 
 $C_{\mathcal{E},1}$   & 0.0005&  0.00025 &  0.000125 \\ \hline
 $C_{\mathcal{E},2}$   & 0.001 & 0.0005 &  0.00025  \\ \hline
  	\end{tabular}
\end{center}
\caption{\giachi{Contact constant $C_c$ and step size near the plate $\Delta t_{plate}$ for the bouncing disk case.}}
\label{Cc}
\end{table}

\begin{figure}[t!]
 \centering
% I)\includegraphics[scale=0.55]{Figures/6e7basket2e6scaling}\\\vspace{0.25cm}
% II)\includegraphics[scale=0.55]{Figures/6e7basket6e10scaling.pdf}
I)\includegraphics[scale=0.55]{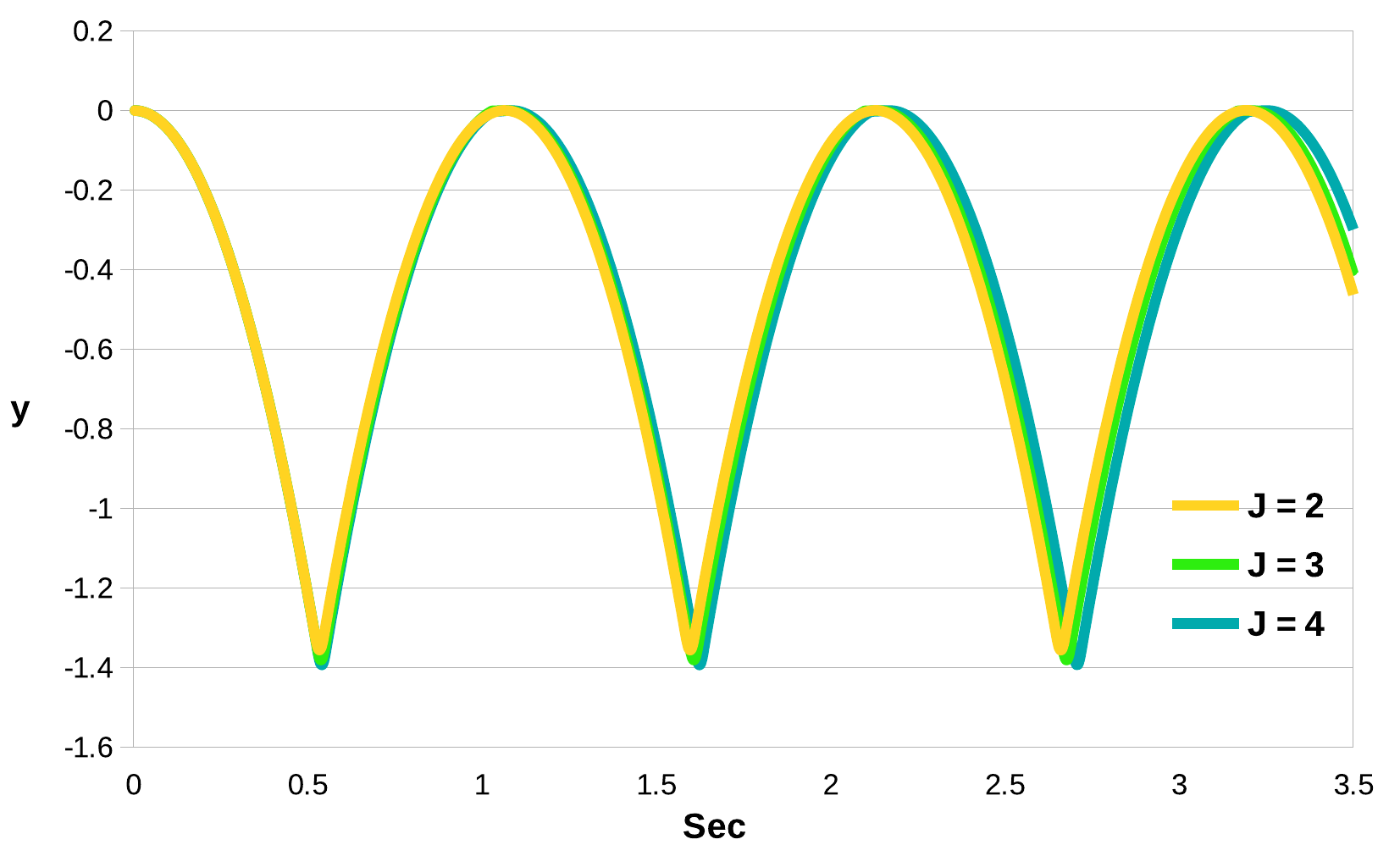}\\\vspace{0.25cm}
II)\includegraphics[scale=0.55]{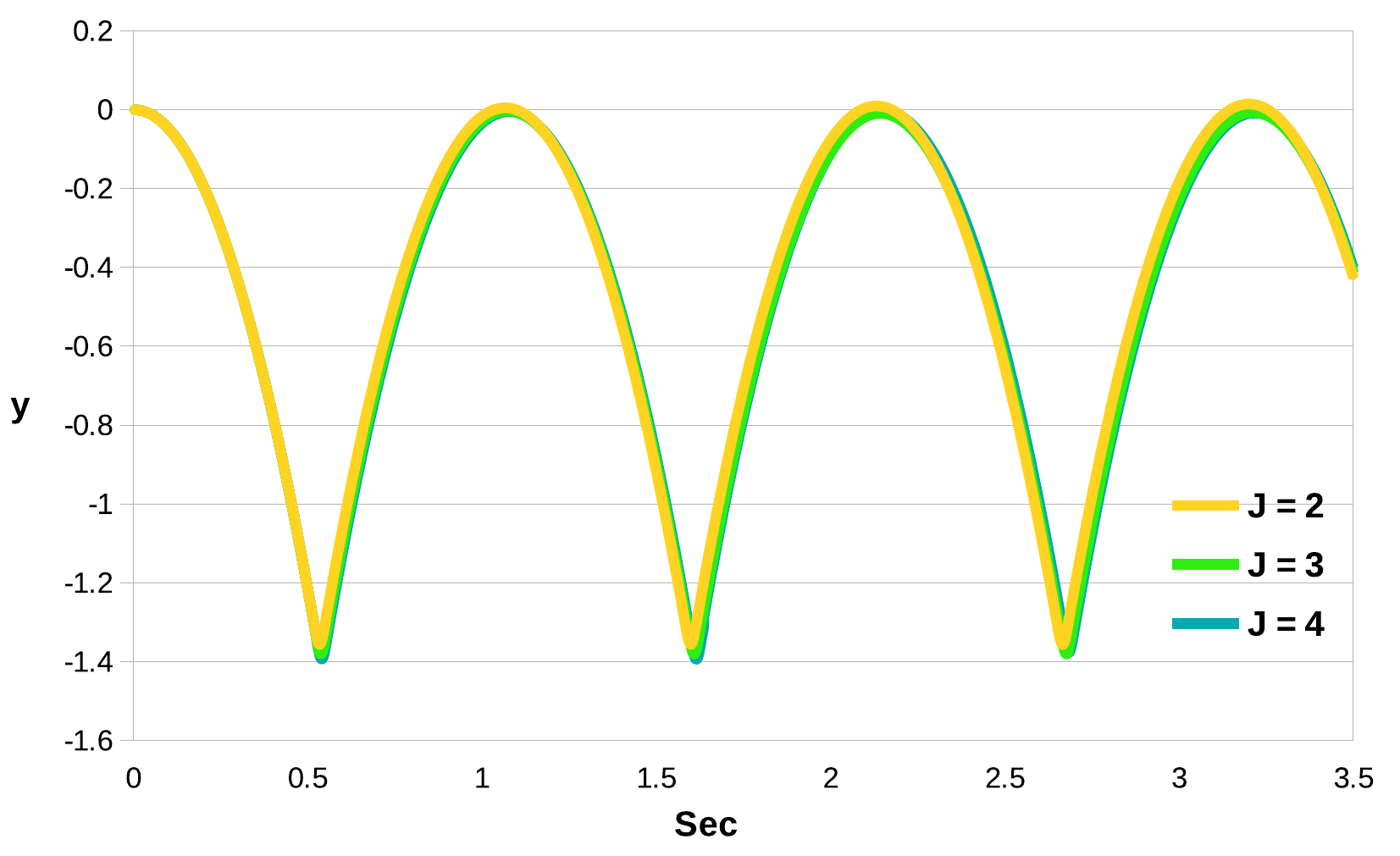}
\caption{Vertical position of the center of mass of a disk bouncing on a plate. I) $C_{\mathcal{E}} = C_{\mathcal{E},1}$ defined in Eq. \eqref{CE3}. II) $C_{\mathcal{E}} = C_{\mathcal{E},2}$ defined in Eq. \eqref{CE4}.}\label{2e6}
\end{figure}

The results for $C_{\mathcal{E},1}$ are displayed in Figure \ref{2e6} I), where the vertical position of the center of mass is tracked as time increases. It is assumed that the reference frame is positioned on the initial position of the center of mass and that the plate is not subject to the gravitational force, to avoid oscillations not caused by the impact.
\giachi{As the refinement of the background grid increases, a slight offset between the curves is observed, the magnitude of which increases with time. An analogous situation was observed in Figure \ref{beamResults}, for the case of an oscillating cantilever beam.
A different situation is displayed in Figure \ref{2e6} II), for $C_{\mathcal{E},2}$.
Compared to Figure \ref{2e6} I), it can be seen that the offset is considerably reduced and now the all the curves are in better agreement, especially those for $J=3$ and $J=4$.
In both cases shown, no damping of the oscillations is observed, once again confirming the reliability and accuracy of our monolithic coupling.}
In conclusion, the results presented here show that, for fixed values of the Young's modulus, it is important for the accuracy to choose the least possible values of $C_{\mathcal{E}}$ that guarantees the convergence of the method.

\section{Conclusion}
A new coupling procedure between the material point method and the finite element method has been presented.
The material point method has been formulated implicitly, and unnecessary exchanges of information between the particles and the background grid have been avoided, resulting in an improved accuracy \giachi{compared to the case where this transfer of information is performed}.
The coupling with the finite element method has been carried out in a monolithic fashion, eliminating the need for time consuming contact search algorithms used by existing strategies to determine the relative position of the bodies.
While numerical results have shown accuracy and reliability of the newly developed monolithic coupling procedure,
further work has to be done to explore its capabilities.
\giachi{For instance, different types of couplings other than solid-solid will be investigated, as well as techniques to use our monolithic approach for the coupling of the MPM with a fluid-structure interaction framework.}
\giachi{
\section*{Appendix A}
\subsection*{Initialization of the MPM disk pseudocode:}
\begin{align*}
   & \mbox{Input Parameters: } x_c,\, y_c ,\, R ,\, R_0,\, N_{\theta_0},\, N_b\, \; \left( \text{with } N_b \ge \frac{R-R_0}{R_0} \text{ ceil}\left(\frac{N_{\theta_{0}}}{2 \pi}\right)\right)\\
   &N_r = \text{ceil}\left(\frac{N_{\theta_{0}}}{2 \pi}\right), \quad \Delta R = \frac{R_0}{N_r} \\
   & \mbox{\#Uniform Distribution}\\
   &p = 1, \quad x_p = x_c, \quad y_p = y_c \\
   &\mbox{\textbf{for} } i = 0,..., N_r - 1\\ 
     &\qquad  r_i = R_0 - i \, \Delta R \,, \quad  N_{\theta_i} = \text{ceil} \left( N_{\theta_0} \, \frac{r_i}{R_0} \right), \quad \Delta \theta_i = \frac{2 \pi} {N_{\theta_i}}\\
     &\qquad \mbox{\textbf{for} } j = 0,..., N_{\theta_i} - 1\\
       &\qquad \qquad p = p + 1, \quad x_p = x_c + r_i \cos(j\,\Delta \theta_i), \quad y_p = y_c  + r_i \sin(j\,\Delta \theta_i)\\
     &\qquad \mbox{\textbf{end for} }\\
   &\mbox{\textbf{end for} }\\
   &\mbox{\textbf{for} } i = 1,..., p\\
     &\qquad m_p = \rho_{mpm} \frac{\pi R_0^2}{p} \\
   &\mbox{\textbf{end for} }\\
   &\mbox{\#Boundary Layers} \\
   &\mbox{\textbf{Find $a$ such that}} \sum_{k=1}^{N_b} \Delta R \, a^k = R - R_0\\
   &\mbox{\textbf{for} } i = 1,..., N_b\\ 
     &\qquad  r_i = R_0 + \sum_{k=1}^{i} \Delta R \, a^k\,,\quad  N_{\theta_i} = \text{ceil}\left( \frac{N_{\theta_0}}{a^i}\right)\,, \quad \Delta \theta_i = \frac{2 \pi} {N_{\theta_i}}\\
     &\qquad \mbox{\textbf{for} } j = 0,..., N_{\theta_i} - 1\\
       &\qquad \qquad p = p + 1, \quad x_p = x_c + r_i \cos(j\,\Delta \theta_i), \quad y_p = y_c  + r_i \sin(j\,\Delta \theta_i)\\
       &\qquad \qquad m_p =  \rho_{mpm} \, r_i \, \Delta \theta_i \, \Delta R \, a^i \\
     &\qquad \mbox{\textbf{end for} }\\
   &\mbox{\textbf{end for} }
\end{align*}
\subsection*{Initialization of the MPM beam pseudocode:}
\begin{align*}
   & \text{Input Parameters: } x_0,\, y_0 ,\, H, L ,\, H_0,\, N_H,\, N_b \; \left( \text{with } N_b \ge \frac{H-H_0}{2} \frac{N_H-1}{H_0}\right)\\
   &\Delta H = \frac{H_0}{N_H - 1}\,,\quad  L_0 = L - \frac{H - H_0}{2}\,,\quad N_L = \mbox{ceil}\left(\frac{L_0}{\Delta H}\right)\,, \quad \Delta L = \frac{L_0}{N_L-1}\\
   & \mbox{\#Uniform Distribution}\\
   &p = 1 \\
   &\mbox{\textbf{for} } i = 0,..., N_L-1 \\ 
     &\qquad \mbox{\textbf{for} } j = 0,..., N_H-1 \\
       &\qquad \qquad p = p + 1, \quad x_p = x_0 + i \, \Delta L , \quad y_p = y_0 - \frac{H_0}{2}  + j \Delta H\\
       &\qquad \qquad m_p = \rho_{0}\, \Delta L \, \Delta H\\
     &\qquad \mbox{\textbf{end for} }\\
   &\mbox{\textbf{end for} }\\
   &\mbox{\#Boundary Layers} \\
   &\mbox{\textbf{Find $a$ such that}} \sum_{k=1}^{N_b} \Delta H \, a^k = \frac{H - H_0}{2}\\
   &\mbox{\textbf{for} } i = 1,..., N_b\\ 
   &\qquad H_i = H_0 + 2 \sum_{k=1}^{i} \Delta H \, a^k \,, \quad N_{H_i} = \text{ceil} \left(\frac{H_i}{\Delta H}\right) + 1 \, , \quad \Delta H_i = \frac{H_i}{N_{H_i} - 1} \\
   &\qquad L_i = L_0 +   \sum_{k=1}^{i} \Delta H \, a^k \,, \quad N_{L_i} = \text{ceil} \left(\frac{L_i}{\Delta H}\right) + 1 \, , \quad \Delta L_i = \frac{L_i}{N_{L_i} - 1} \\
     &\qquad \mbox{\textbf{for} } j = 0,..., N_{L_i} - 1\\
       &\qquad \qquad p = p + 1\,, \quad x_p = x_0 + j \Delta L_i\,, \quad y_p = y_0  + \frac{H_0}{2}\\
       &\qquad \qquad m_p =  \rho_{0}\, \Delta L_i \, \Delta H_i\\
     &\qquad \mbox{\textbf{end for} }\\
     &\qquad \mbox{\textbf{for} } j = 1,..., N_{H_i} - 1\\
       &\qquad \qquad p = p + 1\,, \quad x_p = x_0 + L_i\,, \quad y_p = y_0  + \frac{H_0}{2} -j \Delta H_i\\
       &\qquad \qquad m_p =  \rho_{0}\, \Delta L_i \, \Delta H_i\\
     &\qquad \mbox{\textbf{end for} }\\
     &\qquad \mbox{\textbf{for} } j = 1,..., N_{L_i} - 1\\
       &\qquad \qquad p = p + 1\,, \quad x_p = x_0 +L_i - j \Delta L_i\,, \quad y_p = y_0  - \frac{H_0}{2}\\
       &\qquad \qquad m_p =  \rho_{0}\, \Delta L_i \, \Delta H_i\\
     &\qquad \mbox{\textbf{end for} }\\     
   &\mbox{\textbf{end for} }
\end{align*}
}

\bibliographystyle{plain}
\bibliography{mpm_fem}

\begin{thebibliography}{10}

\bibitem{femus-web-page}
Eugenio Aulisa, Simone Bn{\'a}, and Giorgio Bornia.
\newblock {FEM}u{S} {W}eb page.
\newblock https://github.com/eaulisa/MyFEMuS, 2017.

\bibitem{aulisa2018monolithic}
Eugenio Aulisa, Simone Bna, and Giorgio Bornia.
\newblock A monolithic {ALE} {N}ewton--{K}rylov solver with
  {M}ultigrid-{R}ichardson--{S}chwarz preconditioning for incompressible
  {F}luid-{S}tructure {I}nteraction.
\newblock {\em Computers \& Fluids}, 174:213--228, 2018.

\bibitem{aulisa2017fluid}
Eugenio Aulisa, Giorgio Bornia, and Sara Calandrini.
\newblock {F}luid-{S}tructure {I}nteraction {M}odeling of {A}rtery {A}neurysms
  with {S}teady-{S}tate {C}onfigurations.
\newblock In {\em VII International Conference on Computational Methods for
  Coupled Problems in Science and Engineering, COUPLED PROBLEMS}, pages
  616--627, 2017.

\bibitem{aulisa2017fluid2}
Eugenio Aulisa, Giorgio Bornia, and Sara Calandrini.
\newblock Fluid-structure simulations and benchmarking of artery aneurysms
  under pulsatile blood flow.
\newblock In {\em 6th ECCOMAS Conference - COMPDYN 2017}, pages 955--974, 2017.

\bibitem{bardenhagen2000material}
SG~Bardenhagen, JU~Brackbill, and Deborah Sulsky.
\newblock The material-point method for granular materials.
\newblock {\em Computer methods in applied mechanics and engineering},
  187(3-4):529--541, 2000.

\bibitem{beuth2008large}
Lars Beuth, Thomas Benz, Pieter~A Vermeer, and Zdzislaw Wikeckowski.
\newblock Large deformation analysis using a quasi-static material point
  method.
\newblock {\em Journal of Theoretical and Applied Mechanics}, 2008.

\bibitem{brackbill1986flip}
JU~Brackbill and HM~Ruppel.
\newblock {FLIP}: A method for adaptively zoned, particle-in-cell calculations
  of fluid flows in two dimensions.
\newblock {\em Journal of Computational Physics}, 65(2):314--343, 1986.

\bibitem{brenner2007mathematical}
Susanne Brenner and Ridgway Scott.
\newblock {\em The mathematical theory of finite element methods}, volume~15.
\newblock Springer Science \& Business Media, 2007.

\bibitem{calandrini2018valve}
Sara Calandrini and Eugenio Aulisa.
\newblock Fluid-structure interaction simulations of venous valves: A
  monolithic ale method for large structural displacements.
\newblock {\em International Journal for Numerical Methods in Biomedical
  Engineering}, 0(0):e3156, 2018.
\newblock e3156 cnm.3156.

\bibitem{calandrini2017magnetic}
Sara Calandrini, Giacomo Capodaglio, and Eugenio Aulisa.
\newblock Magnetic drug targeting simulations in blood flows with
  fluid-structure interaction.
\newblock {\em International journal for numerical methods in biomedical
  engineering}, 34:e2954, 2018.

\bibitem{capodaglio2017particle}
Giacomo Capodaglio and Eugenio Aulisa.
\newblock A particle tracking algorithm for parallel finite element
  applications.
\newblock {\em Computers \& Fluids}, 159:338--355, 2017.

\bibitem{charlton2017igimp}
TJ~Charlton, WM~Coombs, and CE~Augarde.
\newblock i{GIMP}: An implicit generalised interpolation material point method
  for large deformations.
\newblock {\em Computers \& Structures}, 190:108--125, 2017.

\bibitem{chen2015improved}
ZP~Chen, XM~Qiu, X~Zhang, and YP~Lian.
\newblock Improved coupling of finite element method with material point method
  based on a particle-to-surface contact algorithm.
\newblock {\em Computer Methods in Applied Mechanics and Engineering},
  293:1--19, 2015.

\bibitem{Ciarlet:2002:FEM:581834}
Philippe~G. Ciarlet.
\newblock {\em Finite Element Method for Elliptic Problems}.
\newblock Society for Industrial and Applied Mathematics, Philadelphia, PA,
  USA, 2002.

\bibitem{cummins2002implicit}
SJ~Cummins and JU~Brackbill.
\newblock An implicit particle-in-cell method for granular materials.
\newblock {\em Journal of Computational Physics}, 180(2):506--548, 2002.

\bibitem{gilmanov2008hybrid}
Anvar Gilmanov and Sumanta Acharya.
\newblock A hybrid immersed boundary and material point method for simulating
  {3D} fluid--structure interaction problems.
\newblock {\em International journal for numerical methods in fluids},
  56(12):2151--2177, 2008.

\bibitem{guilkey2003implicit}
James~Edward Guilkey and Jeffrey~A Weiss.
\newblock Implicit time integration for the material point method: Quantitative
  and algorithmic comparisons with the finite element method.
\newblock {\em International Journal for Numerical Methods in Engineering},
  57(9):1323--1338, 2003.

\bibitem{guilkey2007eulerian}
JE~Guilkey, TB~Harman, and B~Banerjee.
\newblock An {E}ulerian--{L}agrangian approach for simulating explosions of
  energetic devices.
\newblock {\em Computers \& structures}, 85(11-14):660--674, 2007.

\bibitem{guo2006three}
YJ~Guo and JA~Nairn.
\newblock Three-dimensional dynamic fracture analysis using the material point
  method.
\newblock {\em Computer Modeling in Engineering and Sciences}, 16(3):141, 2006.

\bibitem{hamad2017interaction}
Fursan Hamad, Zdzislaw Wikeckowski, and Christian Moormann.
\newblock Interaction of fluid--solid--geomembrane by the material point
  method.
\newblock {\em Computers and Geotechnics}, 81:112--124, 2017.

\bibitem{hogan2014fast}
Robin~J Hogan.
\newblock Fast reverse-mode automatic differentiation using expression
  templates in {C}++.
\newblock {\em ACM Transactions on Mathematical Software (TOMS)}, 40(4):26,
  2014.

\bibitem{ionescu2005computational}
Irina Ionescu, James Guilkey, Martin Berzins, Robert~M Kirby, and Jeffrey
  Weiss.
\newblock Computational simulation of penetrating trauma in biological soft
  tissues using the material point method.
\newblock {\em Studies in health technology and informatics}, 111:213--218,
  2005.

\bibitem{lian2012coupling}
Yanping Lian, Xiong Zhang, and Yan Liu.
\newblock Coupling between finite element method and material point method for
  problems with extreme deformation.
\newblock {\em Theoretical and Applied Mechanics Letters}, 2(2), 2012.

\bibitem{lian2011coupling}
YP~Lian, X~Zhang, and Y~Liu.
\newblock Coupling of finite element method with material point method by local
  multi-mesh contact method.
\newblock {\em Computer Methods in Applied Mechanics and Engineering},
  200(47-48):3482--3494, 2011.

\bibitem{lian2012adaptive}
YP~Lian, X~Zhang, and Y~Liu.
\newblock An adaptive finite element material point method and its application
  in extreme deformation problems.
\newblock {\em Computer methods in applied mechanics and engineering},
  241:275--285, 2012.

\bibitem{love2006energy}
E~Love and DL~Sulsky.
\newblock An energy-consistent material-point method for dynamic finite
  deformation plasticity.
\newblock {\em International Journal for Numerical Methods in Engineering},
  65(10):1608--1638, 2006.

\bibitem{love2006unconditionally}
E~Love and DL~Sulsky.
\newblock An unconditionally stable, energy--momentum consistent implementation
  of the material-point method.
\newblock {\em Computer Methods in Applied Mechanics and Engineering},
  195(33-36):3903--3925, 2006.

\bibitem{ogden1997non}
Raymond~W Ogden.
\newblock {\em Non-linear elastic deformations}.
\newblock Courier Corporation, 1997.

\bibitem{ram2015material}
Daniel Ram, Theodore Gast, Chenfanfu Jiang, Craig Schroeder, Alexey Stomakhin,
  Joseph Teran, and Pirouz Kavehpour.
\newblock A material point method for viscoelastic fluids, foams and sponges.
\newblock In {\em Proceedings of the 14th ACM SIGGRAPH/Eurographics Symposium
  on Computer Animation}, pages 157--163. ACM, 2015.

\bibitem{stomakhin2013material}
Alexey Stomakhin, Craig Schroeder, Lawrence Chai, Joseph Teran, and Andrew
  Selle.
\newblock A material point method for snow simulation.
\newblock {\em ACM Transactions on Graphics (TOG)}, 32(4):102, 2013.

\bibitem{stomakhin2014augmented}
Alexey Stomakhin, Craig Schroeder, Chenfanfu Jiang, Lawrence Chai, Joseph
  Teran, and Andrew Selle.
\newblock Augmented {MPM} for phase-change and varied materials.
\newblock {\em ACM Transactions on Graphics (TOG)}, 33(4):138, 2014.

\bibitem{sulsky2004implicit}
D~Sulsky and A~Kaul.
\newblock Implicit dynamics in the material-point method.
\newblock {\em Computer Methods in Applied Mechanics and Engineering},
  193(12-14):1137--1170, 2004.

\bibitem{sulsky1994particle}
Deborah Sulsky, Zhen Chen, and Howard~L Schreyer.
\newblock A particle method for history-dependent materials.
\newblock {\em Computer methods in applied mechanics and engineering},
  118(1-2):179--196, 1994.

\bibitem{sulsky1996axisymmetric}
Deborah Sulsky and Howard~L Schreyer.
\newblock Axisymmetric form of the material point method with applications to
  upsetting and taylor impact problems.
\newblock {\em Computer Methods in Applied Mechanics and Engineering},
  139(1-4):409--429, 1996.

\bibitem{sulsky1995application}
Deborah Sulsky, Shi-Jian Zhou, and Howard~L Schreyer.
\newblock Application of a particle-in-cell method to solid mechanics.
\newblock {\em Computer physics communications}, 87(1-2):236--252, 1995.

\bibitem{tran2015anisotropic}
Han~D Tran, Deborah~L Sulsky, and Howard~L Schreyer.
\newblock An anisotropic elastic-decohesive constitutive relation for sea ice.
\newblock {\em International Journal for Numerical and Analytical Methods in
  Geomechanics}, 39(9):988--1013, 2015.

\bibitem{wang2016development}
Bin Wang, Philip~J Vardon, Michael~A Hicks, and Zhen Chen.
\newblock Development of an implicit material point method for geotechnical
  applications.
\newblock {\em Computers and Geotechnics}, 71:159--167, 2016.

\bibitem{wikeckowski1999particle}
Zdzislaw Wikeckowski, Sung-Kie Youn, and Jeoung-Heum Yeon.
\newblock A particle-in-cell solution to the silo discharging problem.
\newblock {\em International journal for numerical methods in engineering},
  45(9):1203--1225, 1999.

\bibitem{york2000fluid}
Allen~R York, Deborah Sulsky, and Howard~L Schreyer.
\newblock Fluid--membrane interaction based on the material point method.
\newblock {\em International Journal for Numerical Methods in Engineering},
  48(6):901--924, 2000.

\bibitem{zhang2017incompressible}
Fan Zhang, Xiong Zhang, Kam~Yim Sze, Yanping Lian, and Yan Liu.
\newblock Incompressible material point method for free surface flow.
\newblock {\em Journal of Computational Physics}, 330:92--110, 2017.

\bibitem{zhang2016material}
Xiong Zhang, Zhen Chen, and Yan Liu.
\newblock {\em The material point method: a continuum-based particle method for
  extreme loading cases}.
\newblock Academic Press, 2016.

\bibitem{zhou1999simulation}
Shijian Zhou, John Stormont, and Zhen Chen.
\newblock Simulation of geomembrane response to settlement in landfills by
  using the material point method.
\newblock {\em International journal for numerical and analytical methods in
  geomechanics}, 23(15):1977--1994, 1999.

\end{thebibliography}

\end{document}